\documentclass[a4paper,11pt]{article}
\pdfoutput=1 %

\usepackage{jinstpub} %

\usepackage{lineno}

\usepackage{graphicx}
\usepackage{dcolumn}
\usepackage{bm}
\usepackage{adjustbox}

\graphicspath{{ps/}}
\usepackage{subfigure}
\usepackage{epstopdf}
\usepackage{amsmath}
\usepackage{lineno,hyperref}
\usepackage{makecell}
\usepackage{multirow}
\usepackage{rotating}
\usepackage{enumitem}
\usepackage{color,soul}
\modulolinenumbers[5]

\title{Design and behaviour of the Large Hadron Collider external beam dumps capable of receiving 539 MJ/dump}

\author[a,b]{J.~Maestre}
\author[a,b]{C.~Torregrosa}
\author[a]{K.~Kershaw}
\author[a]{C.~Bracco}
\author[a]{T.~Coiffet}
\author[a]{M.~Ferrari}
\author[a]{R.~Franqueira~Ximenes}
\author[a]{S.~Gilardoni}
\author[a]{D.~Grenier}
\author[a]{A.~Lechner}
\author[a]{V.~Maire}
\author[a]{J.~M. Martin Ruiz}
\author[a]{E.~Matheson}
\author[a]{N.~Solieri}
\author[a]{A.~Perillo-Marcone}
\author[a]{T.~Polzin}
\author[a,d]{V.~Rizzoglio}
\author[a,c]{D.~Senajova}
\author[a]{C.~Sharp}
\author[a]{M.~Timmins}
\author[a,1]{M.~Calviani\note{Corresponding author.}}

\affiliation[a]{Organisation Européenne pour la Recherche Nucléaire (CERN); 1211 Geneva 23, Switzerland}
\affiliation[b]{Universidad de Granada, Spain}
\affiliation[c]{Imperial College London, United Kingdom}
\affiliation[d]{EBG MedAustron GmbH, Marie Curie-Straße 5, 2700 Wiener Neustadt, Austria}

\emailAdd{marco.calviani@cern.ch}

\date{\today}

\abstract{
Two 6~t beam dumps, made of a graphite core encapsulated in a stainless steel vessel, are used to absorb the energy of the two Large Hadron Collider (LHC) intense proton beams during operation of the accelerator. Operational issues started to appear in 2015 during LHC Run~2 (2014-2018) as a consequence of the progressive increase of the LHC beam kinetic energy, necessitating technical interventions in the highly radioactive areas around the dumps. Nitrogen gas leaks appeared after highly energetic beam impacts and instrumentation measurements indicated an initially unforeseen movement of the dumps. A computer modelling analysis campaign was launched to understand the origin of these issues, including both Monte Carlo simulations to model the proton beam interaction as well as  advanced thermo-mechanical analyses. The main ﬁndings were that the amount of instantaneous energy deposited in the dump vessel leads to a strong dynamic response of the whole dump and high accelerations (above 200~g). Based on these ﬁndings, an upgraded design, including a new support system and beam windows, was implemented to ensure the dumps' compatibility with the more intense beams foreseen during LHC Run~3 (2022-2025) of 539~MJ per beam. In this paper an integral overview of the operational behaviour of the dumps and the upgraded configurations are discussed.}


\arxivnumber{2110.08783} 

\begin{document}
\maketitle

\section{Introduction}
\label{Intro}

The two external dumps of the Large Hadron Collider (LHC), also known as Target Dump External (TDE), are essential devices for LHC operation~\cite{Evans_2008}. Their function is to absorb the energy of the two intense counter-rotating beams that circulate inside the LHC vacuum chambers under all foreseeable circumstances, including abnormal operation conditions of the machine. When required, a series of magnets located at Point-6 of the LHC (referred to hereinafter as kickers) deviates each beam trajectory from the LHC ring towards a dedicated extraction line (one per beam) and spreads it out by generating a spiral-like sweep. Each of these two $\sim700$ m long extraction lines ends in a cavern that contains an LHC dump block, where the beam is finally intercepted~\cite{goddard2003}.

\begin{figure}[htpb]
\begin{center}
\includegraphics[width=0.9\linewidth]{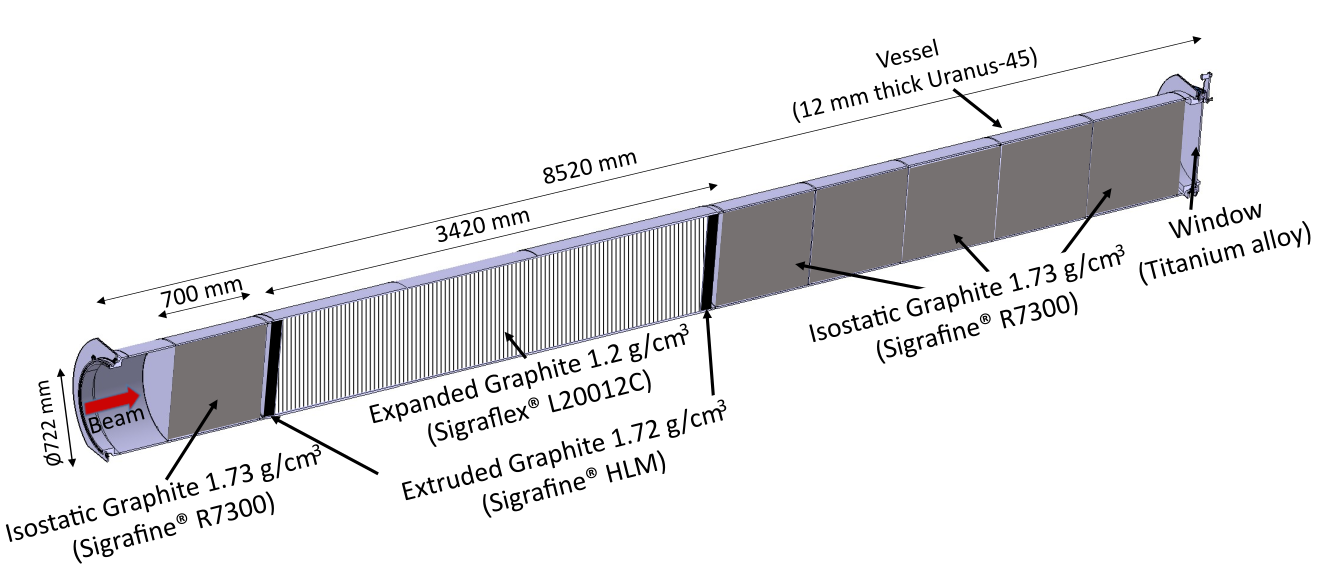}
\end{center}
 \caption{Schematic section of the original LHC dump block showing the graphite core configuration}\label{fig1}
\end{figure}

The original design of the dump block dates from the late 1990s~\cite{Zazula1996nd,Zazula1996er}. It was made up of a series of graphite elements (the core) inside a cylindrical $8520$~mm long $722$~mm diameter duplex stainless steel vessel (318LN, AISI 1.4462, EN 10088-2, also known by its commercial name URANUS\textsuperscript{\textregistered}~45, see Fig.~\ref{fig1}). Different graphite grades, with different densities, were selected to progressively absorb the primary proton beam. The dump vessel was closed at its downstream end by a Titanium window and over-pressurized with nitrogen ($\mathrm{{N_{2}}}$) at $1.2$ bar as a protective atmosphere against graphite oxidation. At its upstream end, the dump block was connected to the LHC extraction line (operated under Ultra-High-Vacuum, UHV) via a connecting line; a UHV beam window was fitted between the UHV extraction line and the nitrogen-filled connecting line. 

The impact of the beam in the LHC dump block generates a particle shower and subsequent deposition of energy in the core and adjacent elements. This phenomenon lasts only $\approx89$ $\mu$s (nominal dilution time), which exposes the dump blocks to high thermo-mechanical loads. Since the start of operation of the LHC in 2009 (Run~1)~\cite{henderson2009}, the beam energy (in each ring of the LHC) has been progressively increased up to  $320$ MJ at the end of Run~2 (2018)~\cite{Steerenberg2019}. Over the course of the coming years, profiting from the upgrades of the LHC Injectors Upgrade (LIU) Project, the beam kinetic energy is expected to reach up to 539~MJ during Run~3 (2022 onwards) in order to maximize the physics reach of the accelerator and its experiments~\cite{Karastathis2019, Rossi2011,schmidt2016high}. These upgrades push the LHC dump blocks to their limits and even beyond their original design~\cite{zazula1997}.

According to the authors’ knowledge, the only similar beam intercepting device that can be compared with the LHC dump block is the Fermilab Tevatron collider external dump~\cite{Kidd1981,Hanna1991}. Albeit designed for a lower beam kinetic energy of $3.2$~MJ, the device incorporates design features that are also present in the LHC beam dump, such as the requirement of being an external dump (i.e. outside of the beam ultra-high-vacuum), the choice of graphite as absorbing material and the requirement of keeping it in an inert atmosphere to avoid graphite oxidation at high temperature. Nevertheless, the energy absorbed by this device was two orders of magnitude below that of the dump block in the LHC. 

Another device featuring a similar beam kinetic energy ($400$ MJ) was the dump for the proposed Superconducting Super Collider (SSC) in the United States~\cite{Cossairt1987}. In the case of the SSC, the design used graphite pellets as the absorber, contained in a water-cooled steel vessel and enclosed in an inert gas atmosphere. A titanium window was envisaged to separate the SSC beam vacuum from the dump core. It is interesting to note that the design considered the graphite as the main absorber material to radially contain the intranuclear cascade shower and to prevent excessive energy deposition on the steel container. However, the SSC project was cancelled during its construction and this device was never studied in detail nor built. 

Certainly, the LHC dump block is a unique device of its kind and there is no previous similar operational experience to learn from. In 2015, after several years of operation without problems (from 2009), some leakage of nitrogen was observed coinciding with the intensity ramp-up of the LHC. Rapid interventions were carried out in the dump caverns to investigate the reasons for the leaks. Initial inspections revealed loose fasteners on the chain clamping rings used to attach the connecting lines to the dump blocks in both caverns.  After removal of the chain clamping rings it was found that the metallic Helicoflex\textsuperscript{\textregistered} gaskets that relied on the chain clamping rings to compress them had been damaged~\cite{maestre2021,martin2021}. Further interventions were carried out covering the replacement of the damaged elements and modification of the nitrogen supply system to maintain the over-pressure even in the event of increased leaks. This ensured the dump block could operate until LHC Long Shutdown 2 which started at the end of 2018. 

\begin{figure}[!htb]
\begin{center}
\subfigure[][]{\includegraphics[width=0.58\linewidth]{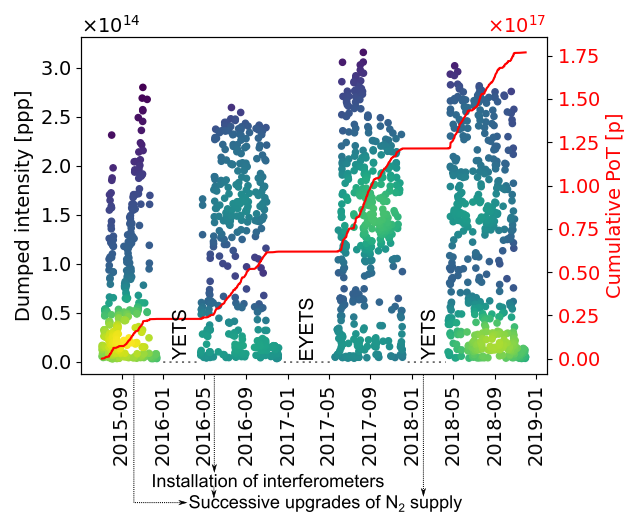}}
\subfigure[][]{\includegraphics[width=0.2\linewidth]{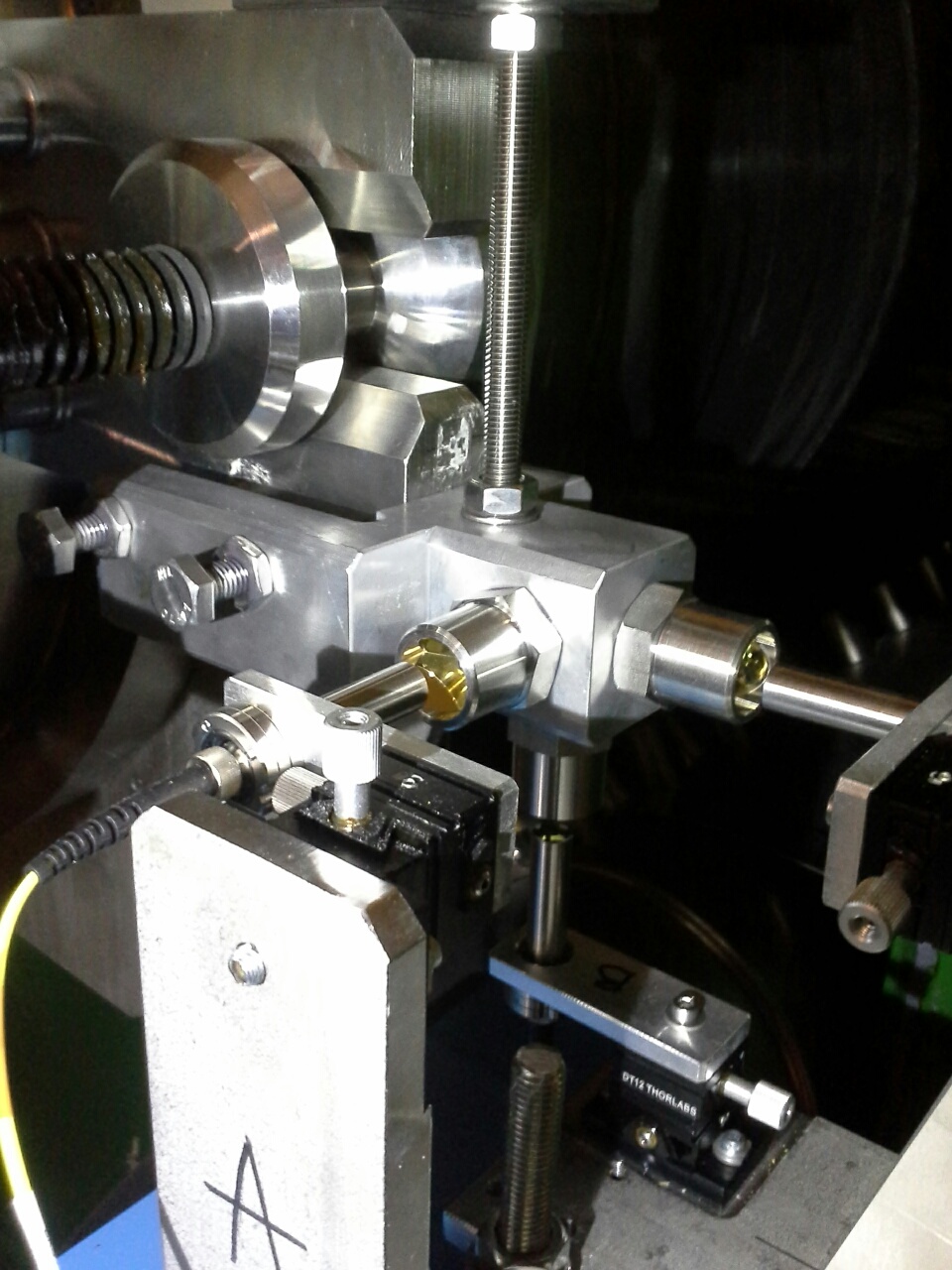}
\includegraphics[width=0.56\linewidth]{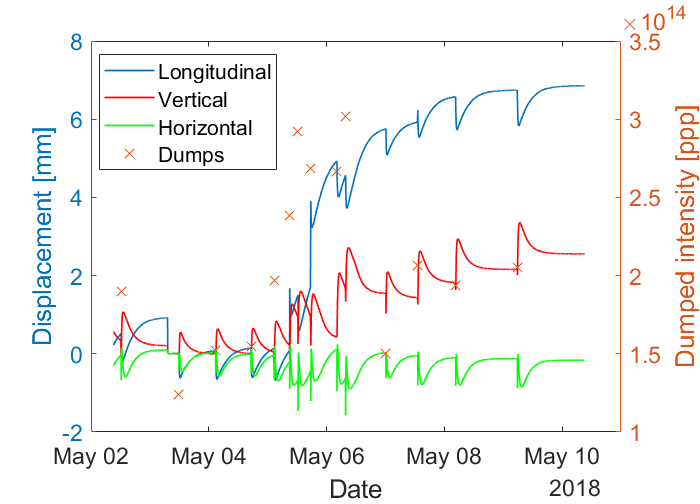}}
\end{center}
 \caption{a) Evolution of dumped beam intensity during Run~2, plotting in dots the total proton intensity per dumped pulse (ppp) and in a continuous line the cumulative number of protons (p) on target/dump (POT). b) Interferometer measurements during a series of dumped beams in May-2018, showing the permanent displacement of the dump block after a series of high intensity dump close to each other in time. The inserted image on the left shows the tri-axial interferometer installed on the dump.}\label{fig2}
\end{figure}

Interventions were made to install passive interferometers to monitor the behaviour of the dump blocks during operation.Fig.~\ref{fig2}a shows the evolution of the dumped beam intensities into the dump block from 2015 to 2019; interventions in the dump caverns were possible during short winter stops, named YETS and EYETS in the figure (Year-End-Technical Stop and Extended Year-End-Technical Stop, respectively). Fig.~\ref{fig2}b shows the displacements recorded by the interferometers over several days in May 2018, together with the intensity of the dumped beam during that period. 

Readings from the instrumentation revealed that the impact of the LHC beam resulted in strong vibrations. In addition, it was observed that the dump blocks had moved longitudinally by several millimeters and also rotated about their longitudinal axes by a few degrees over the course of repeated beam dumps (see Fig. \ref{fig2}b). All these findings, together with the requirements associated with the future LHC exploitation (Run~3), indicated the need for upgrades to be implemented during CERN's Long Shutdown 2 (LS2) in 2019-2021 to make the dump blocks ready for more intense beams~\cite{martin2021}. In order to support these upgrades, complex computer modelling analysis campaigns were launched to improve understanding of the beam dump behaviour. 

Initial findings of the analysis campaigns indicated that the dump block was subjected to extremely high forces and accelerations - and that it would therefore not be possible to prevent damage to the gaskets in the beam line connections by trying to mechanically restrain the dump block. In the original LHC dump configuration, the vibrations of the dump block were transmitted to the connecting line and consequently to the UHV system of the extraction line upstream. Considering the more energetic beam foreseen during Run~3, the risk of damaging the gasket of the vacuum window in the extraction line became a major concern, as it could lead to leaks in the LHC vacuum system. For this reason it was decided to investigate the feasibility of mechanically uncoupling the dump block from the extraction line upstream of the dump. Design work was started to develop the necessary hardware upgrades to achieve this uncoupling. The design work was necessarily an iterative process in parallel with the ongoing analysis campaigns; the proposed changes were repeatedly fed into the computer analysis models until the final designs were arrived at. 

In this paper, these activities are described and discussed. The original configuration of the dump and summary of the upgrades are reviewed in Section \ref{sec:OriginalConf}. In sections~\ref{Fluka},~\ref{mechanical_assesment} and~\ref{dumpblockresults}, a comprehensive numerical analysis of the dump behaviour is presented; FLUKA Monte Carlo was used to model the proton beam interaction with the dump and its output was then fed into thermo-mechanical simulations. Finally, a description and analysis of the upgraded systems, addressing the above mentioned issues, is provided in Sections~\ref{sec:windows} and~\ref{support_design}. 

\section{Original configuration and summary of upgrades during LS2}	 
\label{sec:OriginalConf}

As discussed, each of the two LHC dump blocks is located in a purpose-built cavern forming part of the underground LHC complex. The layout of the cavern before the upgrade work reported here is shown in Fig. \ref{fig2}a. Each dump block is surrounded by a thick iron shielding in order to reduce prompt dose in the surrounding area as well as to partially shield the core residual dose rate during human interventions. During operation, air is continuously flushed underneath the dump core in order to cool it and to circulate the air enclosed between the shielding and dump block. 

In the original configuration, the dump block was physically connected, via a 10 m long connecting line, to the beam extraction line arriving from the LHC ring. An Ultra-High Vacuum (UHV) beam window separated the vacuum in the extraction line from the nitrogen in the connecting line and the dump. The sealing between the incoming extraction line, connecting line, the dump block and its downstream beam window was assured by spring-loaded metallic Helicoflex\textsuperscript{\textregistered}  gaskets between the adjacent ﬂanges of the chambers. These ﬂanges were squeezed against each other by chain clamping rings. These connections were found to be the source of the encountered nitrogen leaks. 

It is important to note that, in its installed position, the dump block was simply resting on two steel supports, with only frictional forces to resist any kind of movement. This freedom is thought to explain the permanent displacements recorded by the instrumentation. It was realised that the observed displacements would have led to pulling and pushing forces on the gaskets and connections with the extraction line. The risk of damaging these connections was a major concern because it could lead to  leaks in the LHC vacuum system. 

\begin{figure}[htpb]
\begin{center}
\subfigure[][]{\includegraphics[width=0.55\linewidth]{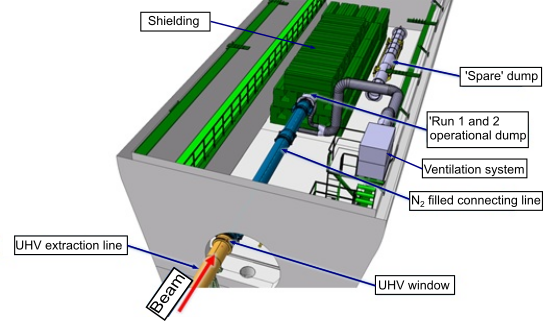}}
\subfigure[][]{\includegraphics[width=0.44\linewidth]{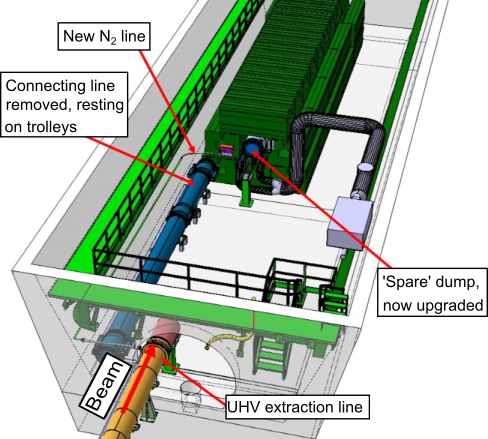}}
\end{center}
 \caption{Integration of the LHC dump block in one of the two beam dump caverns a) before and b) after LS2, showing the main modifications}\label{fig3}
\end{figure}

The main changes to the dump block system implemented during Long Shutdown 2 (2019-2021) in order to safely withstand the kinetic energy increase during LHC Run~3, and that are explained in more detail in the next sections, are summarised below (see Fig. \ref{fig3}b):

\begin{itemize}
    \item Uncoupling of the dump block from the LHC vacuum line to avoid generation of high reaction forces between dump, connecting line and extraction line and consequently to avoid transmission of vibrations to the upstream components.
    \item Design of a new upstream window for the dump block and upgrade of the downstream one. 
    \item Design of a new dump block support system able to cope with the dump block vibrations and prevent long-term physical displacement of the dump.
    \item Installation of a new instrumentation system suite to monitor the dump block behaviour.
\end{itemize}


\section{Beam Characteristics and energy deposition in the dump}
\label{Fluka}

\subsection{LHC beam characteristics}	 
Table~\ref{tab:beamparam} summarizes the proton beam parameters in past and future Runs of the LHC, as well as for the High Luminosity (HL-LHC) upgrade. While the maximum proton energy achieved so far was 6.5~TeV, it will be raised to 6.8~TeV in Run~3. The maximum bunch intensity, which can be handled by different LHC accelerator systems during Run~3, is 1.8$\times$10$^{11}$ protons~\cite{Karastathis2019}. With such bunch intensities, the maximum stored beam energy will increase by almost 70\% compared to Run~2. The LHC beams are produced with two main bunch production schemes, referred to as the standard scheme and the Batch Compression, Merging and Splitting (BCMS) scheme~\cite{Papaphilippou2014}. The latter one, which allows for smaller transverse beam emittances, has been extensively used in Run~2 and is also the baseline scheme for Run~3~\cite{Karastathis2019}. The normalized transverse emittance at top energy is expected to range from 1.8~$\mu$m\,rad to 2.5~$\mu$m\,rad depending on the emittance growth at flat bottom and in the energy ramp~\cite{Karastathis2019}. If not indicated otherwise, all the studies presented in this paper correspond to a bunch intensity of 1.8$\times$10$^{11}$ protons and a transverse emittance of 1.7~$\mu$m\,rad. The latter value is slightly more conservative than the expected emittance values mentioned before. The beam energy was assumed to be 7~TeV in most studies because the actual Run~3 proton energy (6.8~TeV) had not been finally determined when the simulations were carried out.
\begin{table}[htpb]
\centering
\caption{\label{tab:beamparam} Proton beam parameters in the LHC and the HL-LHC era: beam energy $E_{prot}$, bunch spacing $\Delta t_b$, number of bunches per beam $N_b$, bunch intensity $I_b$, stored beam energy $E_{beam}$, and normalized transverse emittance $\varepsilon_n$. The values refer to the beginning of collisions in the experiments. For past runs, peak performance values are reported. In particular, standard 50~ns beam parameters from 2012 are shown for Run~1~\cite{Lamont2013}, while BCMS beam parameters achieved in 2017/2018 are given for Run~2~\cite{Steerenberg2019}. For Run~3, the table lists expected BCMS beam parameters \cite{Karastathis2019}; the specified bunch intensity of 1.8$\times$10$^{11}$ protons might only be reached in the second operational year of Run~3. The last column gives nominal HL-LHC beam parameters for standard 25~ns beams. BCMS beams are considered as an alternative for the HL-LHC era, but are not listed in the table.}
\begin{tabular}{lcccc}
  &Run~1&Run~2&Run~3& HL-LHC\\
&(2009-2013)&(2015-2018)&(2022-2025)&(2028-)\\
\hline
$E_{prot}$ (TeV)& 4 & 6.5 & 6.8 & 7\\
$\Delta t_b$ (ns) & 50 & 25 & 25 & 25 \\
$N_{b}$& 1380 & 2556 & 2748 & 2760\\
$I_{b}$ ($p$)& 1.7$\times$10$^{11}$ & 1.2$\times$10$^{11}$ & 1.8$\times$10$^{11}$ & 2.2$\times$10$^{11}$\\
$E_{beam}$ (MJ) & 150 &  320 & 539 & 680\\
$\varepsilon_n$ ($\mu$m\,rad) & $\approx$2.5 & $\approx$2 & 1.8-2.5 & 2.5\\
\end{tabular}
\end{table}

\subsection{Energy deposition processes, scope and assumptions}	
\label{ssec:flukasim}
The beam dump issues encountered in Run~2 operation were mainly related to the rapid release of energy inside the dump structure. The energy dissipated by high energy proton beams in matter is governed by a variety of physical processes. The copious production of secondary hadrons in inelastic nuclear collisions leads to the build-up of hadronic cascades. The hadronic cascade development is mainly sustained by strongly forward directed pions, protons, neutrons and kaons, and ceases once the secondary particle energy falls below the pion production threshold. Together with the hadronic cascades, electromagnetic cascades are initiated, mainly by photon pairs produced in the decay of neutral pions. The electromagnetic cascades play an important role in dissipating the energy in the dump structure. 
In this section, we present energy deposition estimates for Run~3 beam parameters. All studies were carried out with the FLUKA Monte Carlo code~\cite{Battistoni2015,Bohlen2014,FlukaWeb}. The calculations were based on the assumption that the beam extraction and dilution is correctly executed, i.e. possible failures of the extraction system and their effect on the beam dump are not discussed.

\subsection{Energy absorption and leakage}
\label{ssec:leakage}

The dump absorber blocks, vessel and windows absorb about 78\% of the beam energy, while the remaining energy fraction escapes in the form of secondary particles (19\%) or is consumed in nuclear collisions (3\%). The particles leaking from the dump deposit their energy in the surrounding shielding, air, cavern walls and surrounding bedrock, with the exception of neutrinos, which have a very low interaction cross section and can travel much longer distances. Of all other secondary particles, muons have the largest range in the surrounding bedrock, which can amount to up to a few kilometers. Table~\ref{tab:edep} summarizes the energy deposition in the dump structure and the surrounding environment for Run~3 peak performance parameters. 

\begin{table}[htpb]
\centering
\caption{\label{tab:edep} Energy deposition in the dump and the surrounding environment for projected peak performance parameters in Run 3 (stored beam energy of 539~MJ). The absolute energy deposition values correspond to an emergency beam abort early in a fill. The remaining beam energy is carried away by neutrinos or is spent in nuclear interactions.}
\begin{tabular}{lcc}
     & Fraction of  & Energy deposition\\
     & beam energy & (MJ)\\
\hline
Dump:&&\\
~~~~Graphite &73.6\,\% & 397\\
~~~~Shell  & 4.2\,\% & 23\\
~~~~Windows, flanges etc.  & 0.04\,\% & 0.2\\
\hline
~~~~Total & 77.9\,\% & 420\\
\hline
Environment:&&\\
~~~~Shielding & 17.4\,\% & 94\\
~~~~Air & 0.015\,\% & 0.08\\
~~~~Cavern & 0.12\,\% & 0.6\\
~~~~Molasse, rock, etc. & 0.04\,\% & 0.2\\
\hline
~~~~Total & 17.6\,\% & 95\\
\end{tabular}
\end{table}

The numbers in Table~\ref{tab:edep} correspond to an emergency abort after the beam energy ramping phase, when the stored beam energy in the collider is maximum (539~MJ for a bunch intensity of 1.8$\times$10$^{11}$ protons and a particle energy of 6.8~TeV). In regular physics fills, the dumped intensity will be less due to the burn-off of protons in the experiments and due to proton losses in the LHC collimation system. Premature beam aborts are nonetheless unavoidable and the dump must be able to cope with such scenarios. The energy absorbed by the dump in such a case is expected to be about 420~MJ in Run~3, while it was around 250~MJ in Run~2. Most of the energy is deposited in the massive graphite absorber blocks, which contain the developing particle showers. However, the load on the vessel is still significant due to lateral leakage of particles, in particular around the low-density graphite segment. The total load on the stainless steel vessel is estimated to be 23~MJ in Run~3, whereas it was about 13~MJ in Run~2.   

Most of the energy escaping from the dump is dissipated in the shielding, which consists of concrete-filled dipole yokes from the decommissioned Intersecting Storage Ring (ISR). The energy absorbed by the shielding is expected to be around 100~MJ in Run~3.

\subsection{Transverse beam dilution and energy distribution}

The horizontal and vertical $\beta$-functions at the location of the beam dump are about 5~km and 3.7~km, respectively. For a transverse beam emittance of 1.8~$\mu$m (2.5~$\mu$m), this yields a beam spot size of $\sigma_x \times \sigma_y$ = 1.1~mm $\times$ 1.0~mm (1.3~mm $\times$ 1.1~mm) at a beam energy of 6.8~TeV (without contribution of the dispersion). Even the most robust absorber materials could not sustain the impact of the full beam on a single spot without being damaged. The beam is therefore swept across the dump front face by means of four horizontally and six vertically deflecting kicker magnets~\cite{Bruning2004}. The kicker magnets create phase-shifted sinusoidal fields, which distribute the bunches along a $\sim$1.2~m long ``e''-shaped trace on the dump face. The dilution kickers are located in the LHC tunnel, more than 700~m upstream of the dump cavern. This long distance is needed to achieve the required lateral separation of the bunches when they impact on the beam windows and eventually on the dump core at the end of the transfer line. The beam extraction and dilution is executed seamlessly within one beam revolution, which lasts about 89~$\mu$s.

\begin{figure}[htpb]
\centering
\includegraphics[width=0.5\textwidth]{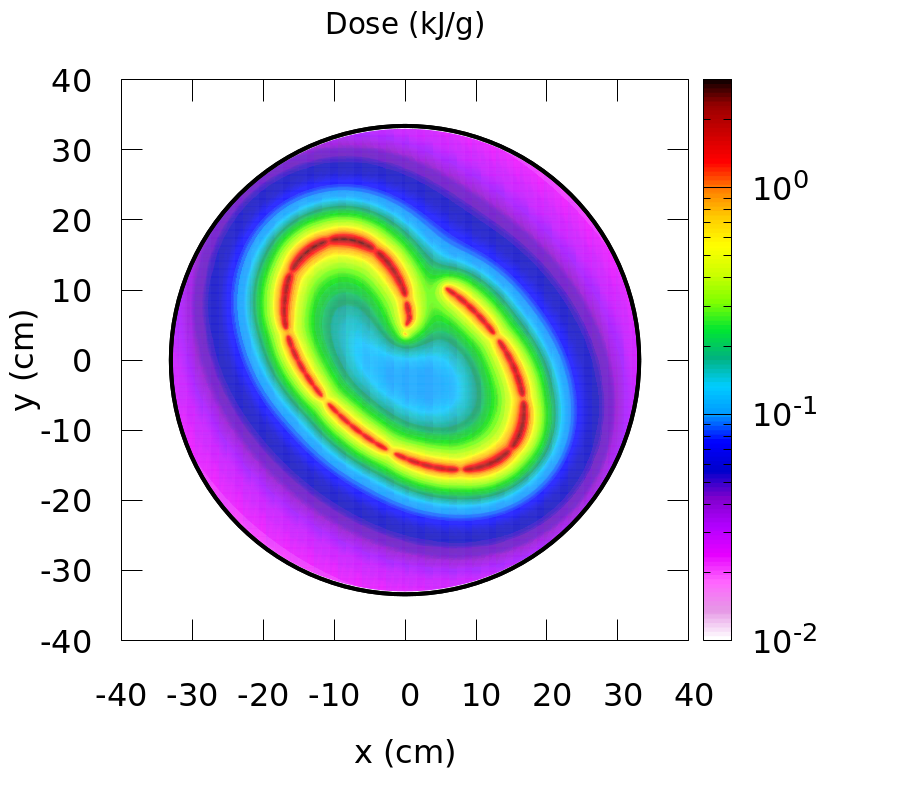}
\caption[minimum]{}
\label{fig4}
Transverse energy deposition density in the low-density graphite segment of the LHC dump core (at a depth of about 270~cm). The results correspond to beam energy of 7~TeV and a bunch intensity of 1.8$\times$10$^{11}$ protons.
\end{figure}

Fig.~\ref{fig4} shows the transverse energy density distribution at a depth of about 270~cm inside the dump core. This corresponds approximately to the location where the shower-induced temperature reaches a maximum. The simulations were carried out for a beam energy of 7~TeV. Slightly lower values can therefore be expected for the actual Run 3 proton energy of 6.8 TeV. For simplicity, the figure neglects any heat diffusion during the sweep. The energy density map illustrates the sweep pattern, which will be adopted in Run~3. The pattern has been slightly modified with respect to Run~2 in order to reduce the voltage in dilution kicker magnets. The energy density and hence the temperature distribution in the core depends on the overlap of the transverse shower tails of the swept proton bunches. The maximum energy density occurs after about 15~$\mu$s, when the vertical field is zero and the sweep velocity is minimum. While individual bunches cannot be discerned in the figure, one can still identify gaps in the bunch filling pattern arising from the beam production in the injectors. These particle-free gaps are needed to accommodate the rise time of the SPS and LHC injection kicker magnets. In the figure, we considered the BCMS beam production scheme, assuming that each LHC injection train from the Super Proton Synchrotron is composed of 240 bunches. The injection trains are arranged in five batches of 48 bunches. The LHC filling scheme might be subject to evolution in Run~3, but the effect on the maximum energy density in the dump core is expected to be small if beam parameters like the bunch intensity and the number of bunches remain similar. FLUKA simulations indicate that the peak energy density might change at most by a few percent if a different filling scheme is used.

\subsection{Longitudinal energy density distribution}

\begin{figure}[htbp]
\centering
\includegraphics[width=0.7\textwidth]{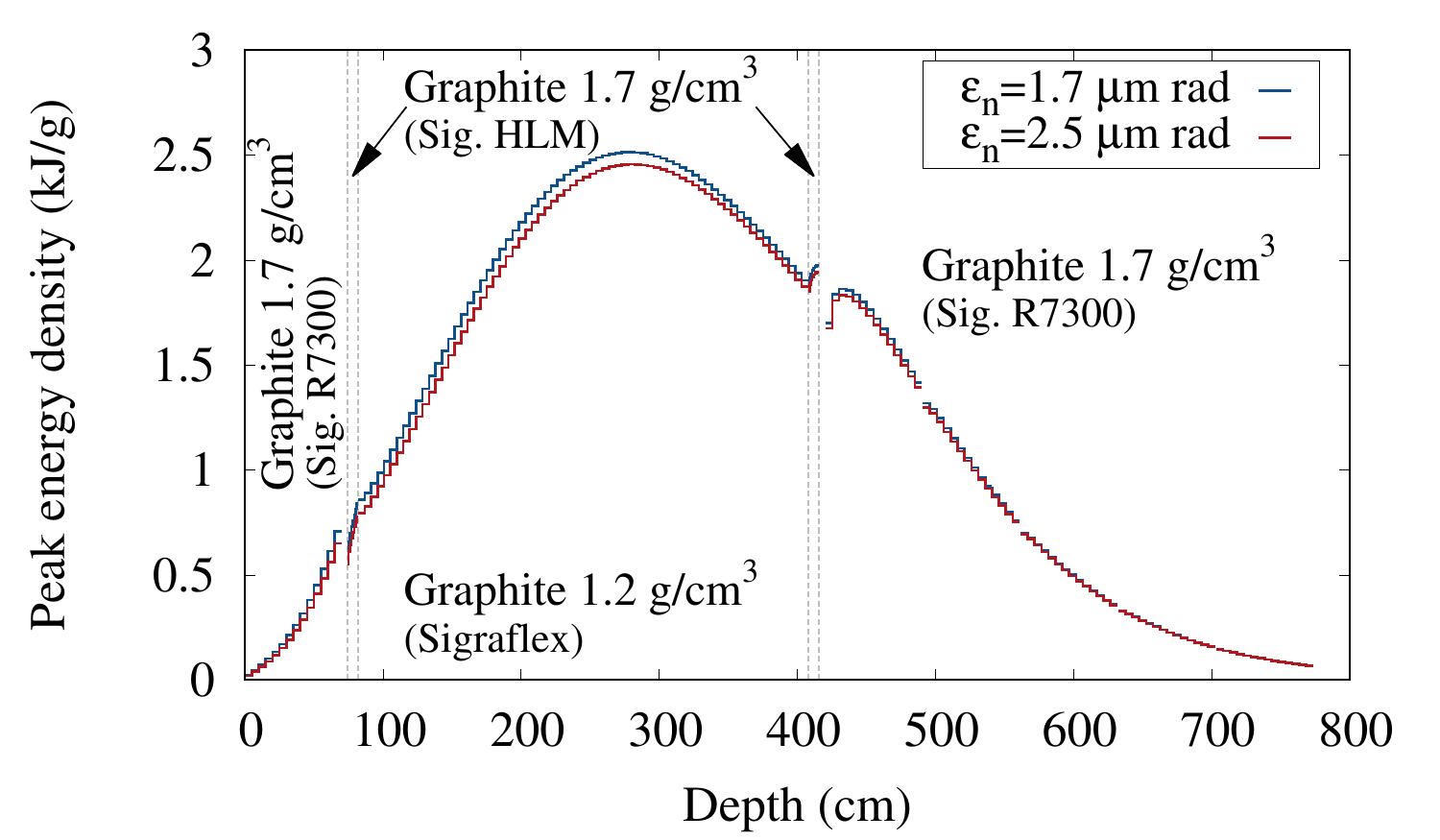}
\caption[minimum]{}
\label{fig5}
Longitudinal peak energy density profile in the dump core. The results correspond to beam energy of 7~TeV and a bunch intensity of 1.8$\times$10$^{11}$ protons. 
\end{figure}

Fig.~\ref{fig5} shows the longitudinal peak energy density profile in the different graphite segments, assuming the same bunch intensity (1.8$\times$10$^{11}$ protons) and filling scheme as before. The beam energy was again considered to be 7~TeV. The shower maximum is reached in the central segment, which is composed of low density Sigraflex\textsuperscript{\textregistered}~ sheets (1.2~g/cm$^3$). 

The lower material density prevents excessive peak temperatures, whereas the higher-density graphite blocks (1.7~g/cm$^3$) allow for a reduced length of the dump while ensuring the required beam absorption. In particular, the higher material density at the upstream end gives rise to a steeper shower build-up, while the higher-density blocks at the downstream end ensure a better attenuation of the longitudinal shower tails. The presence of the upstream higher-density segment has only a small effect on the maximum energy density in the central low-density segment because of the beam dilution. In total, the core represents about 15 inelastic nuclear interaction lengths. This means that the fraction of surviving 7~TeV beam protons is almost negligible (3$\cdot$10$^{-5}$\%) and the energy leakage reported in Section~\ref{ssec:leakage} is largely dominated by secondary particles.

The simulations indicate that the maximum energy density expected in the core with Run~3 bunch intensities is about 2.5~kJ/g, which corresponds to a temperature of about 1400~$^\circ$C. The two curves shown in Fig.~\ref{fig5} correspond to different transverse beam emittances. The two values reflect optimistic and pessimistic emittance values expected in the beginning of stable collisions in Run~3 physics fills. The actual value will depend on the emittance growth rate at flat bottom and during the energy ramp. Although the spot size ($\sigma_x \times \sigma_y$) of individual bunches differs by almost 40\% for the two considered values, the effect on the peak energy density in the dump core is small ($<$3\%). This can be explained by the shower development and the beam sweep. First, the particle showers lead to a transverse dilution of the energy carried by individual bunches. Therefore, the transverse bunch profile is only partially preserved by the transverse energy density profile deep inside the dump. Second, the showers induced by neighbouring bunches in the beam sweep strongly overlap, as illustrated in the previous section. Hence, the peak energy density induced by individual bunches becomes less important. The results demonstrate that the maximum energy density and temperature in the dump is mainly determined by the bunch intensity.

Although the emittance is of little relevance for the temperature rise in the graphite core, it still affects the peak temperature in the upstream dump window since the energy deposited by individual bunches is more focused. We therefore consider the smaller beam emittance (1.7~$\mu$m\,rad) when studying the material response under beam impact in the following sections.

\section{Thermo-mechanical approach of the analyses}
\label{mechanical_assesment}
The impact of the proton beam on the dump block and the ensuing secondary showers results in an energy deposition that produces an almost instantaneous temperature rise. The beam impact duration is only $\mathrm{\sim 89~\mu}$s, hence the initial thermo-mechanical response is highly dynamic. Two different dynamic phenomena may be distinguished and are assessed: firstly the generation of stress waves as a consequence of the material expansion associated with such a fast temperature rise. These waves evolve at the same time as the diluted beam impacts the dump. Subsequently, and after the initial stress wave propagation through the dump block, a general vibration status is developed over a period of a few seconds because of the excitation of several of its natural frequencies. Afterwards, despite being over a longer time scale, the heat transfer phase plays an important role. It redistributes the energy in the dump over time; the energy is finally extracted by radiation and convection to the environment in the dump cavern. This heat transfer phase is a relatively slow process (several hours) and it produces a quasi-static expansion/contraction cycle of the dump block that is also very relevant for the system operation. The analysis and understanding of the full thermo-mechanical behaviour of the dump block during the dynamic and quasi-static responses (referred to as "fast"and "slow" movement, respectively) was essential to address the operational problems described and to design the corresponding upgrades so that the dump could deal with the more energetic Run~3 beam scenarios.

In this section we provide a comprehensive overview of the approach taken for the different analyses performed for the dump block system as a whole. The same basic approach was followed for the beam windows and dump support system - however with some differences.  These differences are explained in sections dedicated to the beam windows and the dump support system.

\subsection{Finite Element Analysis methods}
\label{sec:CompuMethod}
In order to assess the response of the dump blocks after a beam dump event, several Finite Element Analyses (FEA) were carried out. As introduced previously, two different behaviours of the dump blocks can be distinguished: a short intense fast movement in response to the sudden temperature increase that lasts few seconds; and a slow or quasi-static movement during a period of several hours due to the slow thermal energy diffusion. 

Because of the different nature of the aforementioned phenomena and their time frames, different kinds of solvers and software were used to analyse the respective thermo-mechanical behaviour.
Implicit thermo-mechanical solver (ANSYS\textsuperscript{\textregistered}~multi-physics software) was used to analyse the quasi-static movement of the dump block. This technique is able to accurately deal with transient long term simulations thanks to the implicit temporal integration scheme that allows unconditional stable (and adaptive) large time steps. Nevertheless, in fast and short dynamic events, explicit mechanical solvers become advantageous. These are able to cope efficiently with strong non-linear dynamic phenomena with a reduced computational cost. Explicit LS-DYNA\textsuperscript{\textregistered}~software was employed for the analysis of the fast mechanical response of the dump block, weakly coupled with a thermal implicit solver~\cite{liu2008,sun2000}.  

\begin{figure}[htbp]
\begin{center}
\subfigure[][]{\includegraphics[width=0.85\linewidth]{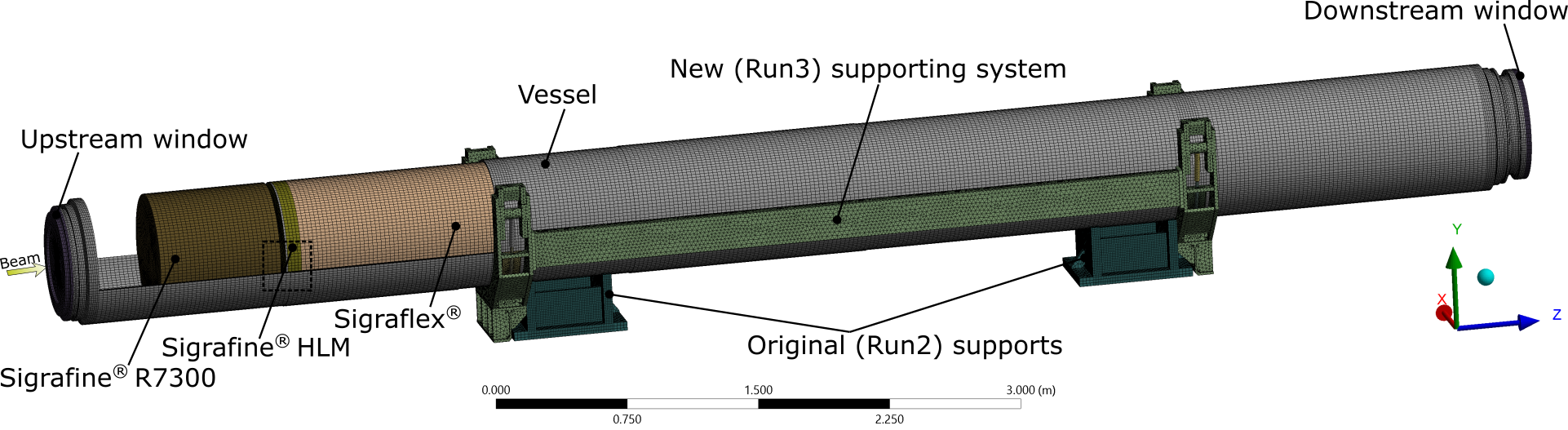}}
\subfigure[][]{\includegraphics[width=0.35\linewidth]{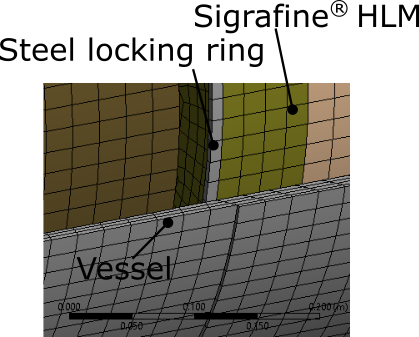}}
\subfigure[][]{\includegraphics[width=0.4\linewidth]{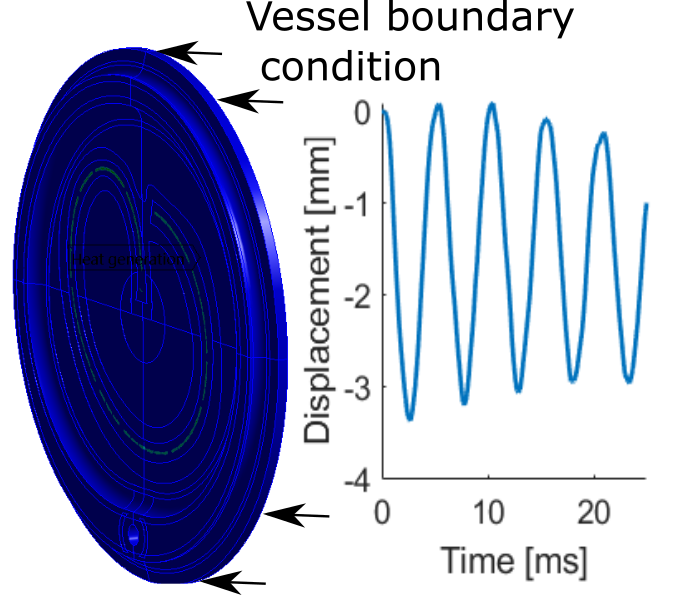}}
\end{center}
\caption{\label{fig6} a) Generic Finite Element Analysis model of the full  dump block. The model is adapted according to the kind of simulation (ranging between $1-2$ millions Degrees of Freedom, DoF) and the mesh is refined around the beam impact area. Although the vessel and windows are relatively thin structures, solid three-dimensional special elements are employed (fully integrated with shear locking reduction algorithm~\cite{borrvall2009}) to capture the energy gradient through the thickness and correctly represent the three dimensional stress status. b) Detail view of the steel locking ring that retains the Sigraflex\textsuperscript{\textregistered}~sector. c) Sub-modeling approach of the upstream windows where the beam is imported as a heat generation load and vessel connection is implicitly considered as a boundary condition.}
\end{figure}

A full FEA model of the dump block system (comprising core, vessel, windows and support system) was developed to study its global response (see Fig.~\ref{fig6}) to a dump event. The dump air cooling system described in Section~\ref{sec:OriginalConf} is implicitly considered as a convective boundary condition. The air-vessel heat transfer coefficient (HTC) was computed via Computational Fluid Dynamic (CFD) simulations and provides a three-dimensional map (as reference, an average HTC of $9$~W/mK on the vessel was computed for an average surface temperature of $75$ $^\circ$C). Radiation heat transfer towards the surrounding (shielding) is also included, assuming a reference external temperature of $22$~$^\circ$C.

The dump block core is made of several graphite grades assembled into the outer vessel following different procedures. The six isostatic Sigrafine\textsuperscript{\textregistered} blocks are shrink-fitted into the vessel (with a nominal interference of $750$~$\mu$m), whilst the low density sector (made of Sigraflex\textsuperscript{\textregistered}~ sheets) is loose-fitted and sits inside the vessel with the only contact forces due to gravity. The Sigraflex\textsuperscript{\textregistered}~sheets are held in place by two HLM Sigrafine\textsuperscript{\textregistered}~plates which are retained longitudinally by two steel locking rings (see Fig. \ref{fig6}b). These details and the pre-stress generated by the shrink-fitting are important in the dynamics of the dump block and are therefore included in the model. Frictional contact is considered between the core and the vessel and a thermal contact conductance (TCC) is computed analytically based on the contact pressure~\cite{madhusudana1996}. Welds and bolted connections are assumed as bonded contacts. The assembly methods are recreated in the model in order to realistically represent the initial pre-stressed status of the  components. 

To analyse in detail the different sub-systems, in particular the beam windows (critical components exposed to direct beam impacts) and the dump block support system, sub-modelling approaches were used. This technique allows to focus the numerical model only on the sub-system and provides a higher level of resolution and accuracy than the full model. The interaction between the systems is directly included in the full model and transferred to the sub-system by imposing displacements as boundary conditions on the common interfaces~\cite{narvydas2014}. Fig.~\ref{fig6}a illustrates this approach for the upstream beam window.   


\subsection{Beam load as input of FEA models}
\begin{figure}[htbp]
\begin{center}
{\includegraphics[width=0.55\linewidth]{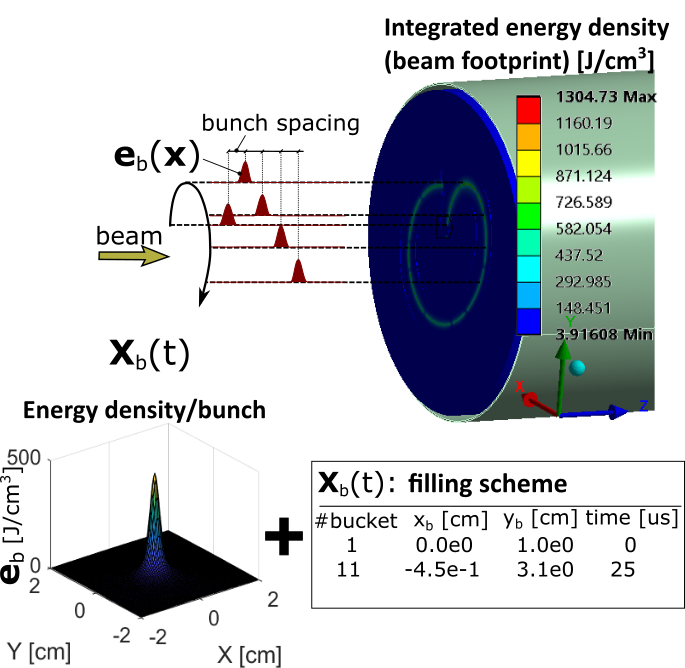}}
\end{center}
\caption{Illustration of the beam load application on the FEA model based on the bunch energy density (generated by FLUKA Mpnte Carlo code) and filling scheme. FLUKA Monte Carlo results are imported using an ad-hoc Python plugin. The FEA mesh is adapted to reduce importation errors (with a global error less than $0.1 \%$).}
\label{fig7}
\end{figure}

Although the beam load is common to all the FEAs related to beam intercepting devices, its application depends on the kind of simulation and purpose. Results from FLUKA Monte Carlo are imported as a distributed heat generation source by an ad-hoc Python plugin (see Fig.~\ref{fig7}). To dynamically recreate  the beam dilution pattern (required for the analysis of the waves' propagation, for example), the proton beam is re-constructed at the level of the bunches' distribution (See Section~\ref{ssec:flukasim}). This tool uses as input the three-dimensional energy density deposition map of a single bunch, together with the filling scheme containing the position and time of each bunch along the dilution path, and it generates (as output) a series of time-dependent heat generation curves for each element of the finite element analysis model, including the time-dependent bunch information. Alternatively when the dynamic bunch structure of the beam is not relevant from the mechanical response point of view (this is the case after the waves' propagation), a step-wise heat source, in which the beam load is integrated over time may be conveniently applied and it substantially reduces the computational cost. 

\subsection{Fatigue assessment method}
\label{fatigue_assesment}
The dump block is subjected to a series of complex vibrations generating multi-axial stresses after each beam dump event. Based on previous Run~2 experience, it was conservatively estimated (and taken as the nominal requirement) that the dump block system should withstand $400$ high energy dumps/year at full intensity over the next $4$ years of Run~3 operation. Fatigue analysis was carried out on the vessel, beam windows (which maintain the enclosed protective environment of the core) and the support system, in order to ensure reliable operation of the dump as a whole.

A wide variety of multi-axial fatigue criteria can be found in the literature, each one proposed for different purposes and/or conditions~\cite{balthazar2007}. In the framework of this study, several fatigue criteria based on the stresses: Sines~\cite{sines1959}, Crossland~\cite{crossland1956}, Dang Van~\cite{van1989} and critical plane: Fatemi-Socie~\cite{fatemi1988} are assessed. An ad-hoc computer aided fatigue analysis approach (FEA-Python code) was specifically developed. The code post-processes the stress and strain history of the FEA and generates a fatigue life map of the full model. Depending on the stress conditions, two different algorithms were  implemented. For cases where the overall principal stress shows a variable pattern, cumulative linear damage based on Palmgren-Miner's rule~\cite{miner1945} and Rainflow counting method~\cite{AMZALLAG1994287} was used, giving the amount of equivalent damage as output. In cases of a quasi-constant stress amplitude, the code looks for the most harmful cycle from the time history and returns the number of cycles up to failure and equivalent amplitude fatigue stress. 
 
\section{Results and discussion of the dump block response to beam impact}
\label{dumpblockresults}
\subsection{Quasi-static behaviour ("slow" response)}
\begin{figure}[htbp]
\begin{center}
\subfigure[][]{\includegraphics[width=0.80\linewidth]{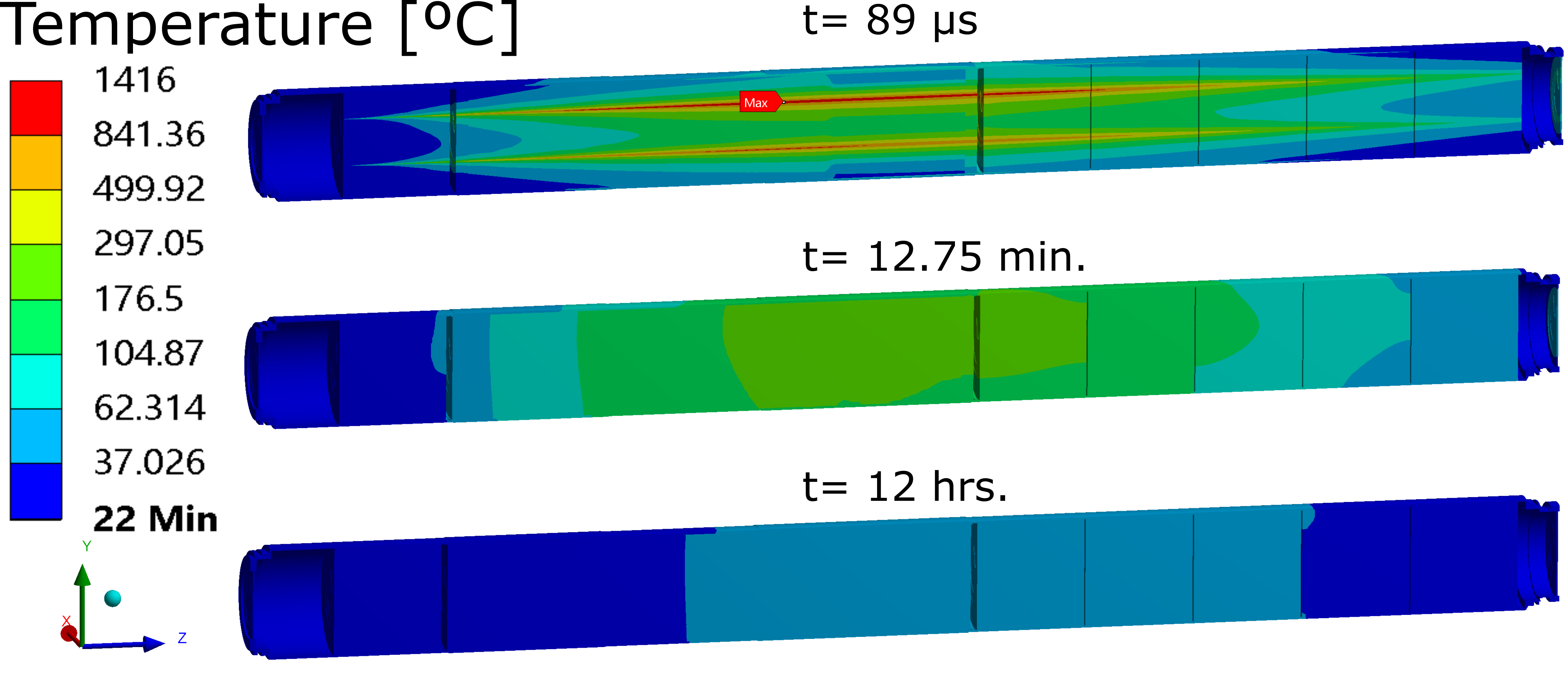}}
\subfigure[][]{\includegraphics[width=0.65\linewidth]{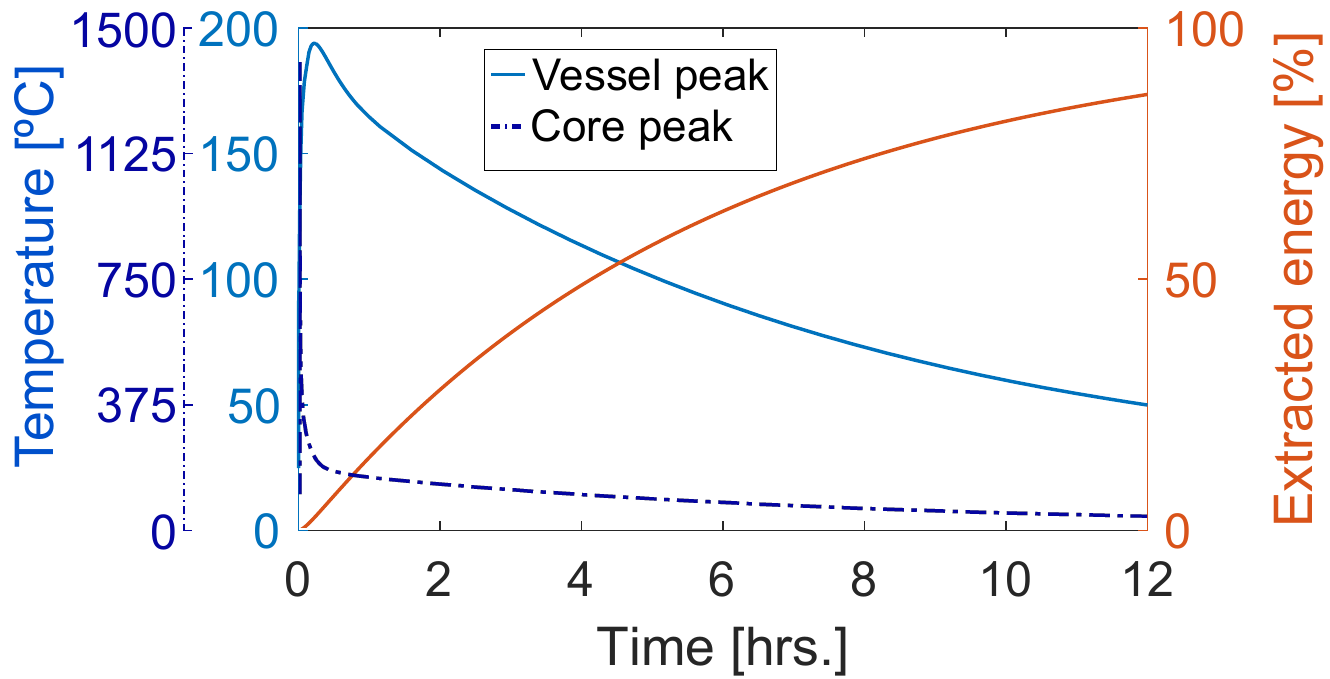}}
\end{center}
\caption{a) Temperature distribution in the dump block (longitudinal section) at different time steps after dumping the beam illustrating the temperature evolution during a cycle of 12 hrs (assuming no beam dump events in between). b) Temperature evolution of the vessel and core along with extracted energy to the surroundings with respect to the total energy deposited in the dump block.}
\label{fig8}
\end{figure}

Fig.~\ref{fig8}a shows the simulated temperature distribution in the dump block at different time steps following a beam dump event. During the beam impact, the temperature increase is maximized around the beam impact region with a maximum temperature of $\mathrm{T_{max}=1416}$ $^{\circ}$C in the Sigraflex\textsuperscript{\textregistered}~sector for nominal Run~3 parameters. The initial high temperature gradient in the core together with the relatively high thermal conductivity of the graphite, and especially the Sigraflex\textsuperscript{\textregistered}~ ($70$ W/Km  and $238$ W/Km at room temperature for the Sigrafine\textsuperscript{\textregistered} and Sigraflex\textsuperscript{\textregistered}, respectively), allows a rapid diffusion of the thermal energy towards the stainless steel vessel. This leads to a temperature rise in the vessel up to a maximum of $\mathrm{T_{max}=194}$~$^{\circ}$C after roughly 13 minutes; this is followed by a progressive cooling down over several hours in which the energy is extracted by convection and radiation to the surroundings (see Fig.~\ref{fig8}b). 

Even after roughly $12$ hours, $13.4\%$ of the deposited thermal energy in the dump block remains inside the core. The low Biot number (dimensionless index that relates the internal conduction with the superficial convection)~\cite{bergman2011}, $Bi=0.04$, explains the quasi uniform radial temperature distribution observed in the core during the cooling down phase. Note that a Biot number far from unity indicates that the system is not thermally well balanced.  A sensitivity analysis carried out on several thermal parameters confirmed that the heat transfer conductivity between air and vessel is the key parameter that limits the average power that can be deposited in the dump core in the long term. Although the cooling time could be potentially reduced by improving the cooling system, it is acceptable for Run~3 (where nominally two beam dumps are assumed per day) in the current configuration. However, the limited cooling capacity of the current dumps will be a bottle neck in the future, especially in the High-Luminosity LHC era.

\begin{figure}[htbp]
\begin{center}
\subfigure[][]{\includegraphics[width=0.75\linewidth]{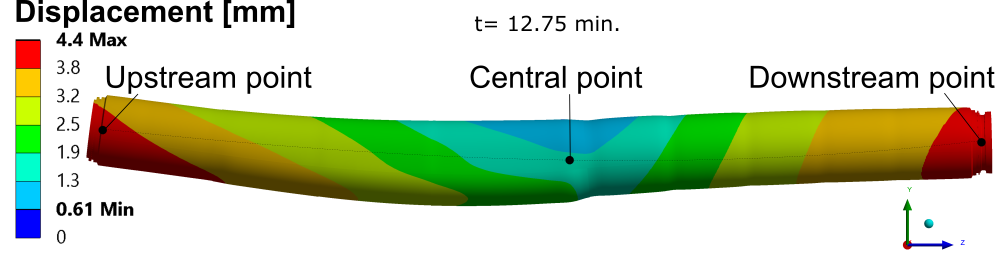}}
\subfigure[][]{\includegraphics[width=0.7\linewidth]{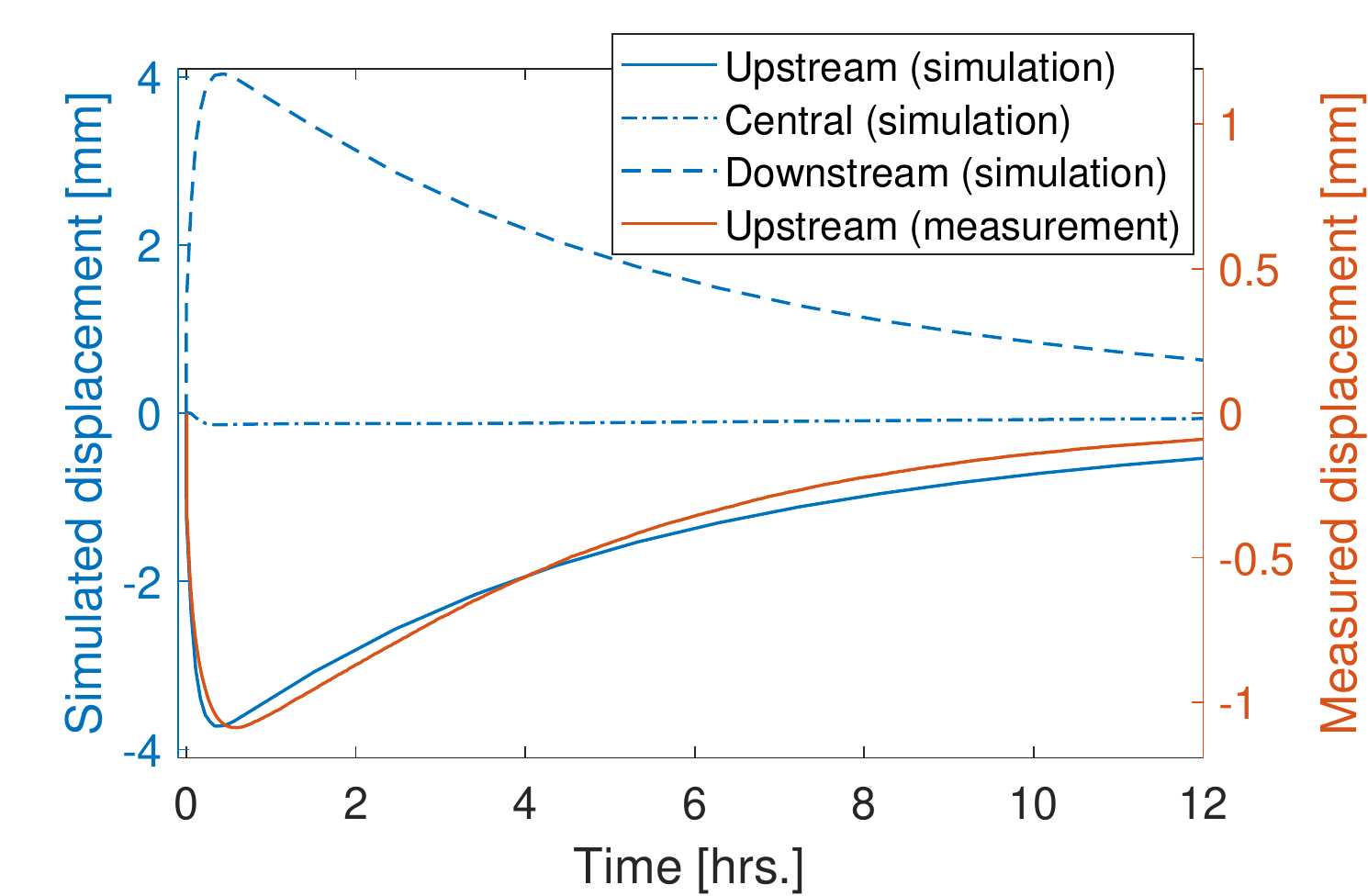}}
\end{center}
\caption{Illustration of the quasi-static mechanical evolution of the dump block following a beam dump event: a) deformation at the time of maximum expansion; and b) the evolution of the longitudinal displacements at different points (upstream flange, $4.2$~m with respect to this flange and downstream flange) over a cycle of 12 hrs (Run~3 nominal intensity, $4.95\cdot10^{14}$ protons at $7$ TeV, is assumed for the simulation). For comparison purposes the displacement measured during Run~2 is included (dated on 9\textsuperscript{th} July 2018 with an effective dumped intensity of $2.05\cdot10^{14}$ protons at $6.5$ TeV). Positive values indicate a displacement in the beam direction.}
\label{fig9}
\end{figure}

The quasi-static mechanical evolution of the dump block is illustrated in Fig.~\ref{fig9}. The dump expands longitudinally  by up to 4~mm over around 40 mins and then contracts, following a cycle similar to the temperature evolution of the vessel. At the same time, it experiences a slight bending deformation due to the non-axisymmetric boundary conditions and core  material distribution (as shown in Fig.~\ref{fig9}a). The reason for this behaviour is to be found in the temperature change of the vessel during the slow heat transfer from the core and subsequent cooling. The high Young's modulus and thermal expansion coefficient of the steel of the vessel (195~GPa, $12.5\cdot 10^{-6}$~1/K, respectively) compared with the ones of the graphite core (10~GPa,  $2.7\cdot 10^{-6}$~1/K) results in the  dump block deformation being mostly determined by the vessel. Measurements of the displacement at the upstream end of the dump block during Run~2 operation confirm this expansion. It is important to note that both simulations and measurements follow  similar trends (see Fig. \ref{fig9}b), although with different magnitudes. This difference is mainly associated with the different beam energy deposition (nominal Run~3 scenario is considered in the calculations) but also with the intrinsic non-linear behaviour of the mechanical system and the attachment to the connecting line (present during the Run~2 measurements), which opposed the dump block displacements.

\subsection{Dynamic behaviour ("fast" response)}
\label{ssec:fast_response}
\begin{figure}[htbp]
\begin{center}
\subfigure[][]{\includegraphics[width=0.75\linewidth]{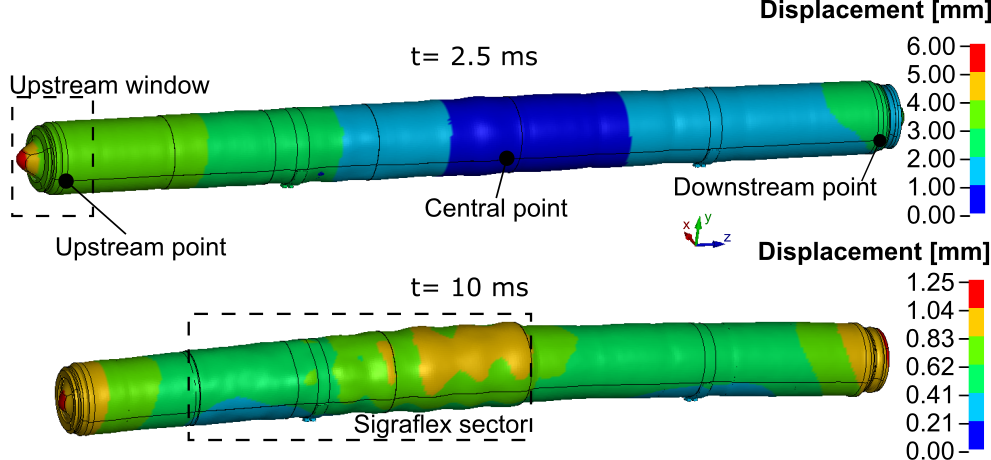}}
\subfigure[][]{\includegraphics[width=0.4\linewidth]{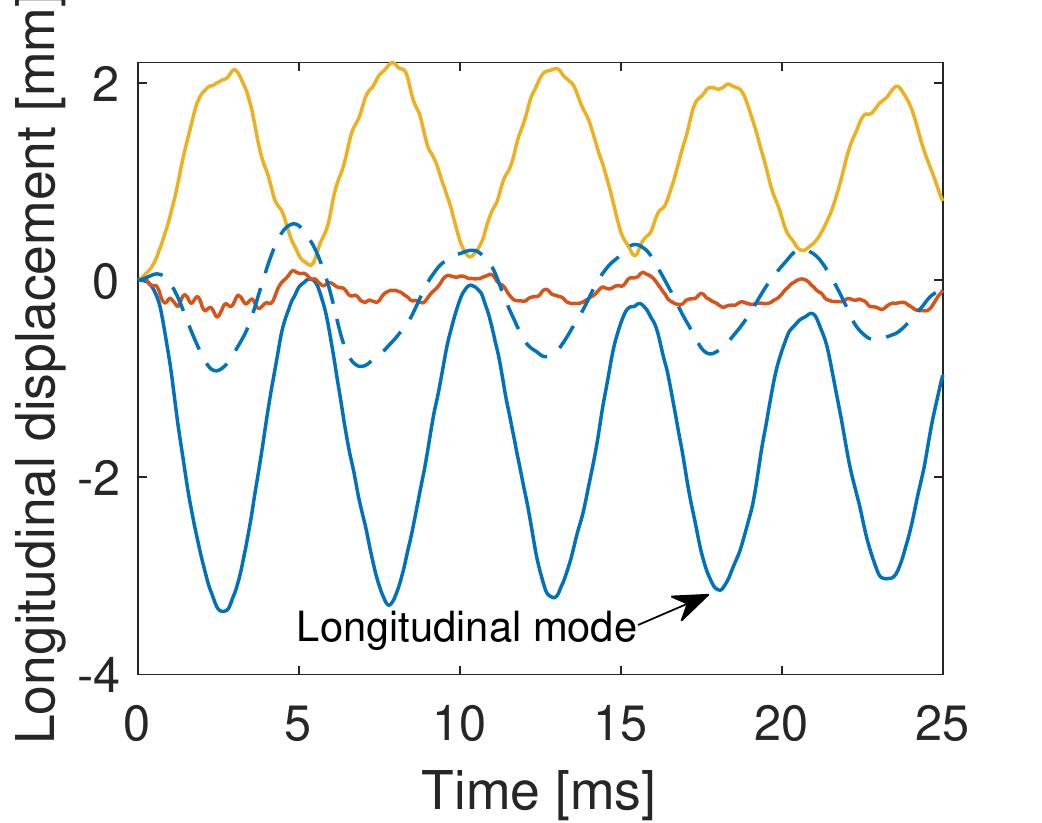}}
\subfigure[][]{\includegraphics[width=0.4\linewidth]{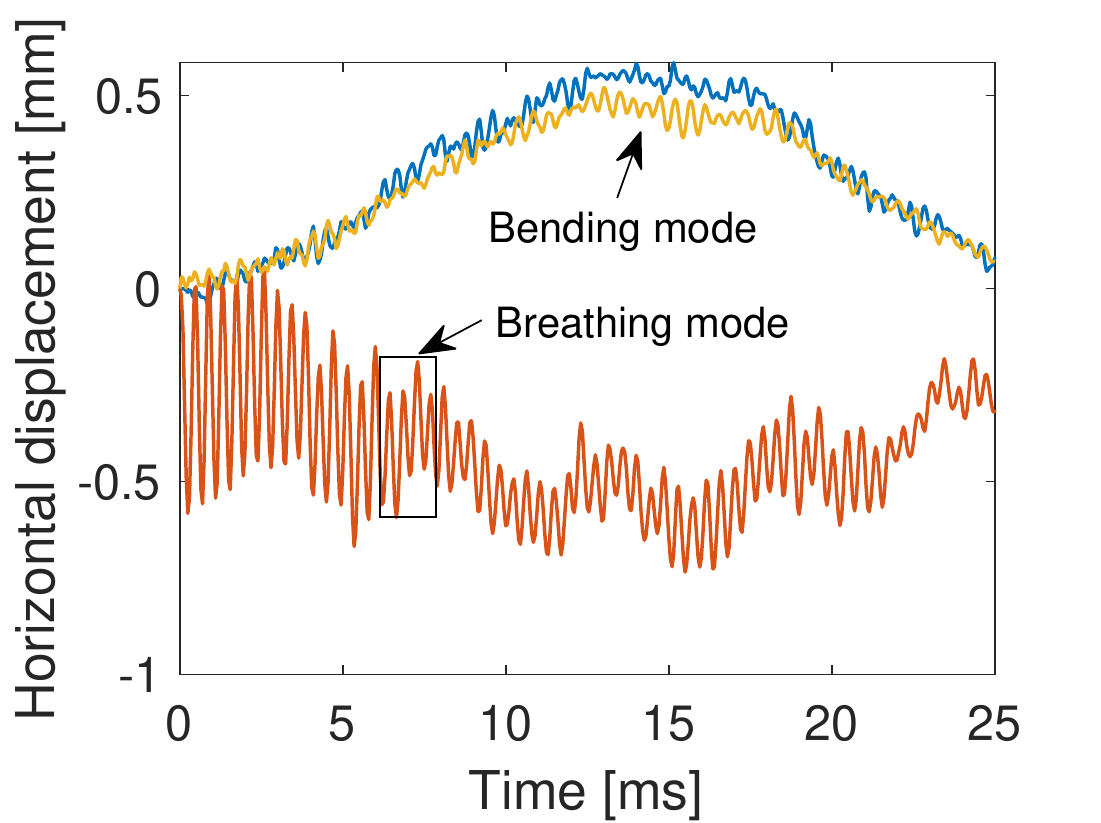}}
{\includegraphics[width=0.3\linewidth]{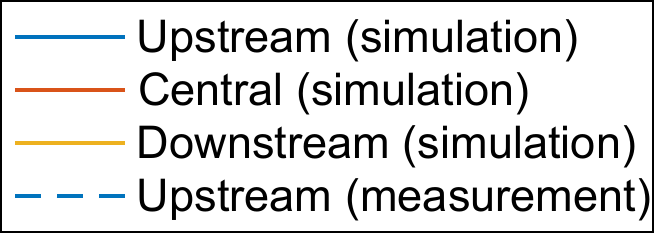}}
\end{center}
\caption{Illustration of the dynamic response of the dump block: a) shows the three main exited natural modes: longitudinal (top); radial and bending (bottom). b) and c) show the longitudinal and horizontal displacement evolution at several locations along the dump block (upstream flange, $4.2$~m with respect to this flange and downstream flange), respectively. For comparison purposes, graph b) includes longitudinal displacement measured during Run~2 operation (measurements from 16\textsuperscript{th} August 2018 with an intensity of $1.04\cdot 10^{14}$ protons at $6.5$ TeV), showing the longitudinal mode in agreement with the simulation but with a lower amplitude.}
\label{fig10}
\end{figure}

Fig.~\ref{fig10} illustrates the dynamics of the dump block displacement a few milliseconds after beam impact. In response to this fast event, after the transmission of mechanical deformation through the whole dump block via wave propagation and subsequent rarefaction, several natural frequencies are excited in the assembly. Unlike the quasi-static behaviour, dominated by the heat diffusion phenomenon, the fast movement is a direct consequence of the sudden energy deposition in the stainless-steel vessel, in which inertia and mass play an important role.

From simulations, it is evident that the longitudinal mode is the main contributor to the dump dynamics. As a consequence of this mode, both windows of the dump block: the original downstream and also the new upstream one (foreseen in the new Run~3 configuration introduced in Section~\ref{sec:OriginalConf}) are shaken by the dump movement with acceleration peaks of $\approx 1900$~m/s\textsuperscript{2} and $1000$~m/s\textsuperscript{2}, respectively (see top of Fig.~\ref{fig10}a and b). This phenomenon has important implications for the design of the new upstream window, as will be described in Section~\ref{sec:windows}. In addition, there are two other relevant modes: the bending and radial modes (see bottom of Fig.~\ref{fig10}a and c). The radial mode is clearly perceptible in the Sigraflex\textsuperscript{\textregistered}~sector, whilst the phenomenon is partly attenuated in the shrink-fitted graphite blocks sector. This radial mode is of particular concern for the support system as the radial expansion induces high forces ($\approx 2 \times$ dump weight) in support components (more details of the support system are given in Section~\ref{support_design}).

%

In addition to the previously described numerical analysis, experimental modal analysis was carried out on one of the spare dump blocks being upgraded. This activity had several aims including pre-characterization of the dump before Run~3 operation, extraction of the damping ratios (required for the fatigue analysis) and cross-checking of the numerical simulations. Classical impact techniques were used such as those reported in Ref.~\cite{artoos2007,tirelli2010,Silva2012,Guinchard2018}. 

Table~\ref{tab1} summarises the results of the modal analysis, including both numerical and experimental results for comparison. It is worth noticing the good agreement between them. The longitudinal mode was already detected during the Run~2 operation, thanks to the installed interferometers, at a frequency of $196$ Hz, (see Fig.~\ref{fig10}b), which nicely coincides with both the numerical and modal analysis results. Radial/breathing mode is also detected at 2.3 kHz and it confirms the numerical predictions. 

\begin{table}[htpb]
\caption{Experimental~\cite{Scislo2020} and numerical modal analyses of the dump. Two numerical analyses are reported assuming a linear modal analysis and a non-linear dynamic analysis in the time domain. The damping ratio is referred to the critical damping~\cite{Scislo2020,artoos2007}}
\begin{adjustbox}{center}
\begin{tabular}{cc|ccc}
\multicolumn{2}{c|}{\textbf{Simulations}}                                                                                                                                                        & \multicolumn{3}{c}{\textbf{Experimental measurements}}                                                                                                                                       \\ \hline
\multicolumn{1}{c|}{\begin{tabular}[c]{@{}c@{}}Linear modal analysis\\ Natural freq. {[}Hz{]}\end{tabular}} & \begin{tabular}[c]{@{}c@{}}Dynamic analysis\\ Natural freq. {[}Hz{]}\end{tabular} & \multicolumn{1}{c|}{\begin{tabular}[c]{@{}c@{}}Natural freq. \\ {[}Hz{]}\end{tabular}} & \multicolumn{1}{c|}{Mode shape} & \begin{tabular}[c]{@{}c@{}}Damping ratio \\  {[}\%{]}\end{tabular} \\ \hline
32.0 & - & 27 & Global bending & 0.2 \\
36.9 & 35.4 & 37 & Global bending & 0.2\\
95.3 & - & 102 & Global bending & 4.2 \\
125.1 & - & 134 & Global bending& 1.7 \\
172.3 & - & 183 & Global bending & 1.4 \\
174.4 & 199.4 & 197 & Longitudinal movement & 1.8  \\
186.4 & - & 205 & Global bending & 0.3\\
- & 2353 & 2360 & Radial movement & 0.2                           
\end{tabular}
\end{adjustbox}
\label{tab1}
\end{table}


These vibrations are presumed to be the main source of the detected issues during Run~2 operation, and are presumed to explain the loosening of the fasteners, damage of the Helicoflex\textsuperscript{\textregistered} gaskets (at the connection between dump and connecting line) and the resulting nitrogen leaks. In the original dump support system configuration, the dump block sat on two simple metallic supports which were contoured to match the outer surface of the dump vessel (see Fig.~\ref{fig6}). There were no elements in the original dump support system to prevent longitudinal or rotational displacements of the dump block apart from frictional forces between the outer vessel and the supports. It therefore appears that the vibrations together with the slow expansion and contraction movements of several mm were responsible for the migration of the dump and permanent displacements found in operation. It is thought that the only thing that finally limited these movements was the connection line linking the dump to the beam tube coming from the LHC.

\subsection{Dump vessel fatigue analysis results}
In view of the highly dynamic conditions to which the vessel is subjected, fatigue analyses were carried out on the vessel following the approach described in Section~\ref{fatigue_assesment} and focusing on Run~3 parameters. Table~\ref{tab4} summarizes the results. It can be observed that all the stress-based fatigue criteria are in agreement in terms of number of cycles up to the fatigue strength ($>5\cdot10^6$ cycles), except the modified Dang Van criterion that predicts 1 million cycles. The Dang Van criterion employs the Von-Mises equivalent stress (whereas the other criteria use the deviatoric) and it penalises more strongly the hydrostatic component. The Fatemi-Socie criterion, based on the critical shear plane, provides a number of cycles one order of magnitude higher. Note that the damping phenomenon attenuates the vibrations of the dump block over time such that only the first $120$ cycles after the beam dump (assuming $0.2\%$ damping ratio as measured in the modal analysis) are relevant from the fatigue point of view. Considering, $1600$ beam dumps as the baseline design requirement over the Run~3 dump block lifetime (see Section~\ref{fatigue_assesment}), no fatigue issues are expected (calculated minimum number of beam dump events up to fatigue failure $\approx8900$).  

\begin{table}[htpb]
\caption{Fatigue assessment of the vessel for different criteria assuming the worst condition in the load history.}
\begin{adjustbox}{center}
\begin{tabular}{l|ll}
\textbf{Criterion} &
  \textbf{\begin{tabular}[c]{@{}l@{}} cycles up to \\ fatigue failure\end{tabular}} &
  \textbf{\begin{tabular}[c]{@{}l@{}}Max. Equivalent\\ stress/strain\end{tabular}} \\ \hline
Crossland &
  $5.20\cdot10^{6}$ &
  Eq. Stress $= 165$ MPa \\
Sines &
  $5.22\cdot10^{6}$ &
  Eq. Stress $= 166$ MPa \\
\begin{tabular}[c]{@{}l@{}}Modified \\ Dang Van\end{tabular} &
  $1.11\cdot10^{6}$ &
  Eq. Stress $= 260$ MPa \\
Fatemi-Socie &
  $>1.7\cdot10^{7}$ &
  \begin{tabular}[c]{@{}l@{}}Shear strain $= 0.13$ $\%$\\ Normal Stress $= 170$ MPa\end{tabular}
\end{tabular}
\end{adjustbox}
\label{tab4}
\end{table}

\subsection{Dump vessel instrumentation}
A new instrumentation system was designed and installed during the LS2 upgrade for several purposes: to monitor the thermo-mechanical performance of the dump block, to cross-check numerical simulations and to detect possible anomalies during Run 3 operation. The vessel was therefore equipped with temperature sensors (PT100), strain gauges, displacement sensors (Linear Variable Differential Transformers (LVDT)), velocity sensors (Laser Doppler Vibrometers (LDV)), and acceleration sensors, as well as imaging and acoustic measurement equipment. Details of the instrumentation are provided in Table~\ref{tab:instrumentation}.

\begin{table}[htpb]
\caption{Instrumentation for the dump vessel, including information on the sensor characteristics and acquisition frequencies.}
\begin{adjustbox}{center}
\begin{tabular}{l|lllll}
\textbf{Measurement} & \textbf{Sensor} &\textbf{Number} & \textbf{Frequency} & \textbf{Range} & \textbf{Accuracy}  \\ \hline
Temperature & PT100 & 16 & 10 Hz & 20 - 250 \textdegree \text{C} & $\pm\, 1 $\textdegree \text{C}\\
Strain & Strain Gauges  &  16 & 1\, kHz & $\pm$ 2000\, $\mu \text{m/m}$ & 5-10 $\mu \text{m/m}$\\
Displacement &  LVDT &  3 & 10 Hz & $\pm$ 20 mm & $\pm$ 0.1 mm\\
Velocity & LDV & 1 & 10\, kHz & 0 - 24 m/s & 0.01 m/s\\
Acceleration &   Optical accelerometers & 3 & 240 kHz & 0 - 2000\,g&$\pm$ 10\,\%\\
Images & HD D-SLR & 1 & 1 photo/hr & 1465 x 1010 mm (FOV) & N/A\\
Sound &  Optical Microphone &  1 & 96 kHz & N/A & N/A\\
\end{tabular}
\end{adjustbox}
\label{tab:instrumentation}
\end{table}

\begin{figure}[htbp]
\centering
\resizebox{0.6\textwidth}{!}{
\includegraphics{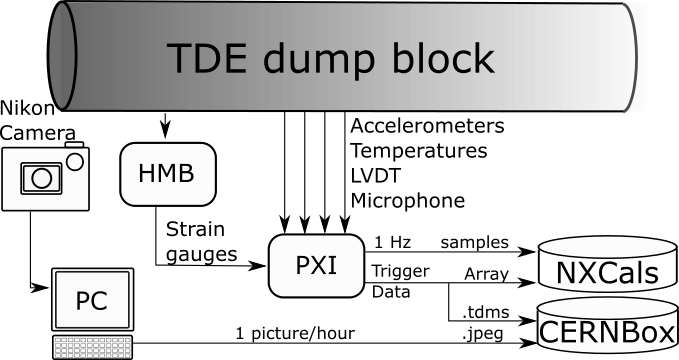}}%
\caption{\label{fig11} General architecture of the instrumentation acquisition. Data are acquired via an NI-PXI system and a Hybrid Multiband (HMB) system then saved in CERN database NXCALS and CERNBox server.}
 \end{figure}

 
All sensors are commercially available, with the exception of the optical accelerometers which have been developed in house at CERN. These are interferometer-based inertial accelerometers with a flexible membrane, that measure acceleration in a single axis. The working principle of this type of accelerometer is to have a plate that linearly deflects with acceleration, while the deformation of all other parts of the device is minimised. The exact accuracy of the system is still under study however the sensors have been calibrated and validated at CERN. The PT100 sensors are from Omega Engineering Ltd, the LDV is provided by OptoMet GmBh, the HD cameras and Nikkor 200-500 lens from Nikon, and the LVDTs from Measurement Specialties (Europe) Ltd. Apart from the camera images which are acquired by a computer (PC), all data will be acquired by a real-time National Instruments (NI) PXI system upon each trigger that indicates a beam dump has taken place. The saved data will consist of a buffer 2 seconds before and 8 seconds after trigger reception. In addition to the the acquisition times referenced in Table~\ref{tab:instrumentation}, a 1 Hz reading from each sensor will be saved to the CERN database NXCALS~\cite{wozniak:icalepcs2019-wepha163} for live viewing. The general software architecture is shown in Figure \ref{fig11}. All acquisition and conditioning electronics, including the high definition cameras, are shielded inside bunkers located in the extraction tunnels, 80~m upstream of the dumps.

%

\section{Design considerations on the TDE dump block windows}
\label{sec:windows}
\subsection{Introduction to window design}

The main function of the dump block beam windows is to provide robust enclosures at the extremities of the stainless steel dump block vessel and contain the internal nitrogen atmosphere while withstanding the beam-induced thermo-mechanical and dynamic loads.

The disconnection of the dump vessel from the upstream vacuum line, carried out during Long Shutdown 2 in order to mitigate the problems induced by vibrations and expansion / retraction cycles, required the addition of a new independent window at the upstream end of the dump block made of Ti-Grade 5 alloy to separate the air of the cavern from the $\mathrm{{N_{2}}}$ atmosphere. Its design is described in this Section.

In addition, a study of the dump block downstream window performance, consisting originally of a 10 mm plate made  from Ti-Grade 2 (unalloyed titanium), indicated that in some Run~3 beam dilution failure scenarios the maximum temperature and stresses reached were  203~$^{\circ}$C  and  209~MPa  respectively, possibly leading to yielding in this material. For this reason, it was also decided to upgrade the design of this window during Long Shutdown 2 by changing it to Ti-Grade 5, with a consequent improvement of the mechanical properties. Due to the poor performance of the original metallic seals and chain clamping rings, it was also decided to change the design of the connection method. This change was implemented by welding a new extension to the dump block with a bolted flange interface, so that the new window could be bolted and sealed via a copper gasket, to the downstream end of the dump vessel. 

\subsection{Design of the new upstream and downstream beam windows}
\begin{figure}[htbp]
\centering
\resizebox{0.35\textwidth}{!}{
\includegraphics{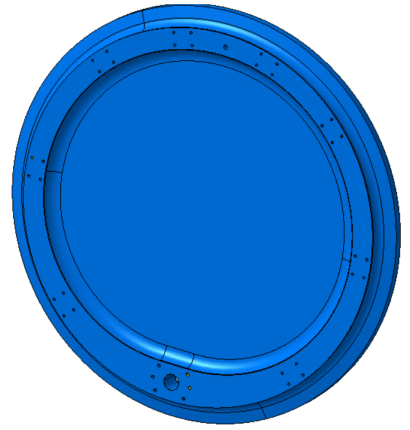}}%
\resizebox{0.35\textwidth}{!}{
\includegraphics{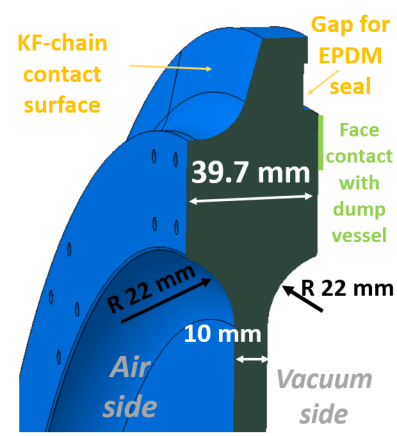}}%
\caption{\label{fig12} Geometry of the new dump upstream window .}
 \end{figure}

Fig.~\ref{fig12} shows front and sectional views of the new dump upstream window. 

The window has a varying thickness; for the central area, the selected thickness is 10~mm. This is the thickness that is directly impacted by the proton beam. Close to the  external radius, the thickness is increased up to nearly 40~mm. This zone of increased thickness – placed at a distance from the beam path of at least $20$~mm – is added to allow the connection of the $\mathrm{{N_{2}}}$ presurization system and to allow bolted attachment of dump monitoring instruments. Furthermore, this increased thickness is vital to provide additional stiffness to the window, to avoid the resonant frequencies of the dump longitudinal vibration, and to minimize the induced dynamic stresses. Finally, a very large transition radius is used between the different thicknesses to minimize dynamic stresses.

\begin{figure}[htbp]
\centering
\resizebox{0.48\textwidth}{!}{
\includegraphics{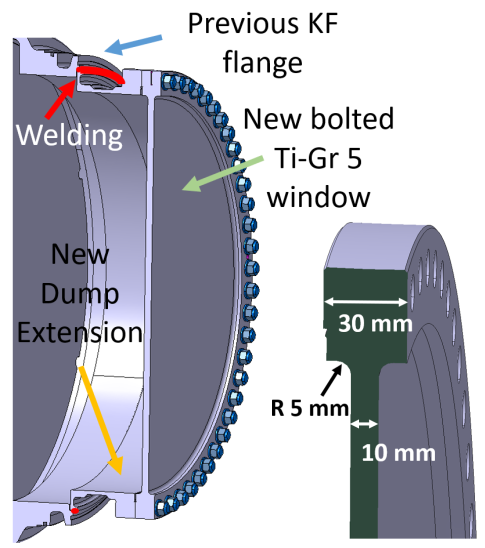}}%
\caption{\label{fig13} Geometry of the upgraded dump downstream window. This new window is bolted, with a CF flange, to an extension which is welded to the vessel.}
 \end{figure}

Fig.~\ref{fig13} shows the area at the downstream end of the dump block and specifically the new configuration of the welded extension and window. The selected thickness of the window is 10~mm, as for the upstream one. However, this window does not have a varying thickness in the same way as the upstream one. This is due to the broader beam footprint in the window, induced by the particle showering inside the core. 

Titanium alloy Grade 5 (also known as Ti-6Al-4V) was selected as material for both the upstream and downstream dump windows. This alloy combines low density (necessary to minimize the beam deposited energy) with exceptionally high yield strength even at high temperatures. The geometry of the windows was machined at CERN by computer numerical control turning and milling machines out of a multi-directional forged disk of 690 mm diameter by 50 mm thickness. Mechanical tests of representative material were carried out according to EN ISO 6892, showing a yield and ultimate strength of 935 and 986 MPa respectively, with associated longitudinal and transversal strain to rupture of 14\% and 39\%. 

The new dump windows must withstand three different types of loads, which take place in different time regimes and are induced by various dynamic sources:
\begin{enumerate}[label=(\roman*)]
\item Thermal stresses directly induced by the energy deposition of the beam during each beam dump event. This energy deposition takes place in $\mathrm{\sim 89~\mu s}$, consistent with the sweeping path arising from the beam dilution system. The window must withstand not only the nominal operation path, but any foreseeable impacting path arising from the dilution kickers failure scenarios. 
\item Mechanical loads induced by the $\mathrm{{N_{2}}}$ atmosphere at 0.2 bars of over-pressure inside the dump block vessel, as well as the loads induced by the internal vacuum required before the nitrogen filling procedure.
\item Dynamic vibration loads arising from the dump block longitudinal vibrations induced by the energy deposition of the beam shower in the vessel (estimated at roughly $\approx200$~Hz, as explained in Section~\ref{ssec:fast_response}). This dynamic load, transmitted by the vessel to the outer circumference of the window, induces additional vibration loads on the window (see Fig. \ref{fig10}). It takes place over a time scale of few hundred milliseconds considering the damping ratio measurements for the longitudinal mode reported in table \ref{tab1}. As will be shown in the results section, the oscillatory stresses on the windows due to this vibration load are the highest ones amongst the three types of loads listed. For this reason, a fatigue analysis using this load was also included in the scope of this work. 
\end{enumerate}

\subsection{Beam induced load analysis and results}
The interaction of the proton beam with the window's material leads to a fast deposition of energy, subsequent temperature rise and dynamic thermo-mechanical stresses. These loads have been analysed using FLUKA Monte Carlo and LS-DYNA\textsuperscript{\textregistered} codes, following the computational approach described in Section~\ref{sec:CompuMethod}. 

\begin{figure}[htbp]
\begin{center}
\subfigure[]{\includegraphics[width=0.45\linewidth]{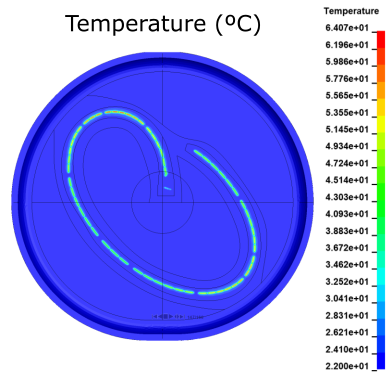}}
\subfigure[]{\includegraphics[width=0.44\linewidth]{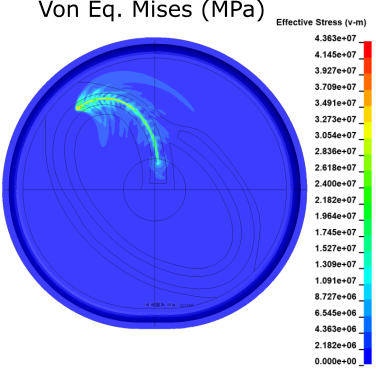}}
\end{center}
 \caption{a) Temperature profile in the dump upstream window at the end of the beam sweep. b) Von Mises dynamic stresses at upstream windows at the time stamp of 17 $\mathrm{\mu}$s. }\label{fig14}
\end{figure}

Fig.~\ref{fig14} shows the contour plots for the temperature and the associated Von Mises equivalent stress obtained by this coupled analysis for the upstream dump window. The temperature plot on the left corresponds to the time stamp at 89~$\mathrm{\mu}$s, once the beam pulse has completed the whole sweep. The maximum temperature reached during the beam-upstream window interaction under nominal conditions is 64~$^{\circ}$C. The stress plot on the right corresponds to the time stamp at 17 $\mathrm{\mu}$s, when the maximum stresses of the beam-induced load is reached ($43$~MPa). This plot also shows very clearly the tails of stress waves that are generated by the beam impact following the sweep pattern. Therefore, in addition to the inertia effects, this analysis takes into account the potential constructive interference of such stress wave fronts.

\begin{figure}[htbp]
\begin{center}
\subfigure[]{\includegraphics[width=0.44\linewidth]{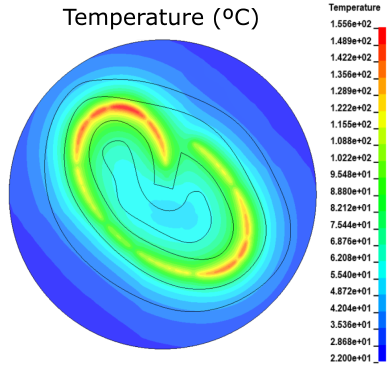}}
\subfigure[]{\includegraphics[width=0.45\linewidth]{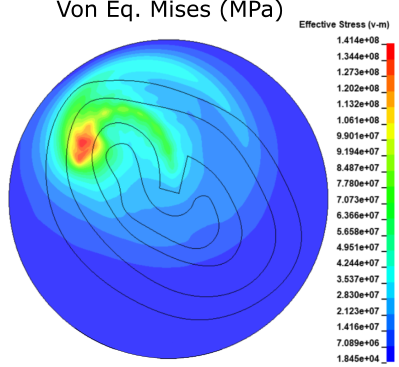}}
\end{center}
 \caption{a) Temperature profile in the dump downstream window at the end of the beam sweep. b) Von Mises dynamic stresses at downstream windows at the time stamp of 32 $\mathrm{\mu s}$}\label{fig15}
\end{figure}

Similarly, Fig.~\ref{fig15}a shows the results for the dump downstream window. The significantly larger area impacted by temperature rise along the beam sweep in comparison to the dump's upstream window is evident.  This is due to the developed particle shower generated along the 8 meters long dump core. The maximum temperature reached in a nominal beam dilution is estimated at 155~$^{\circ}$C. The stress plot shown in figure~\ref{fig15}b corresponds to the time stamp at 32~$\mathrm{\mu}$s, when the maximum stresses of the beam-induced load are reached (corresponding to $141$~ MPa). 

The described analysis demonstrates that the windows are fully compliant with the stresses directly induced by the beam-matter interaction in the windows, since yield strength of Ti-Grade 5 at 150~$^{\circ}$C is at least 650 MPa~\cite{Majidi2008}. This analysis has also been carried out for the most likely beam dilution failure cases, demonstrating compliance also in such cases.

\subsection{Vibration load analysis and results}
\begin{figure}[htbp]
\begin{center}
\subfigure[]{\includegraphics[width=0.49\linewidth]{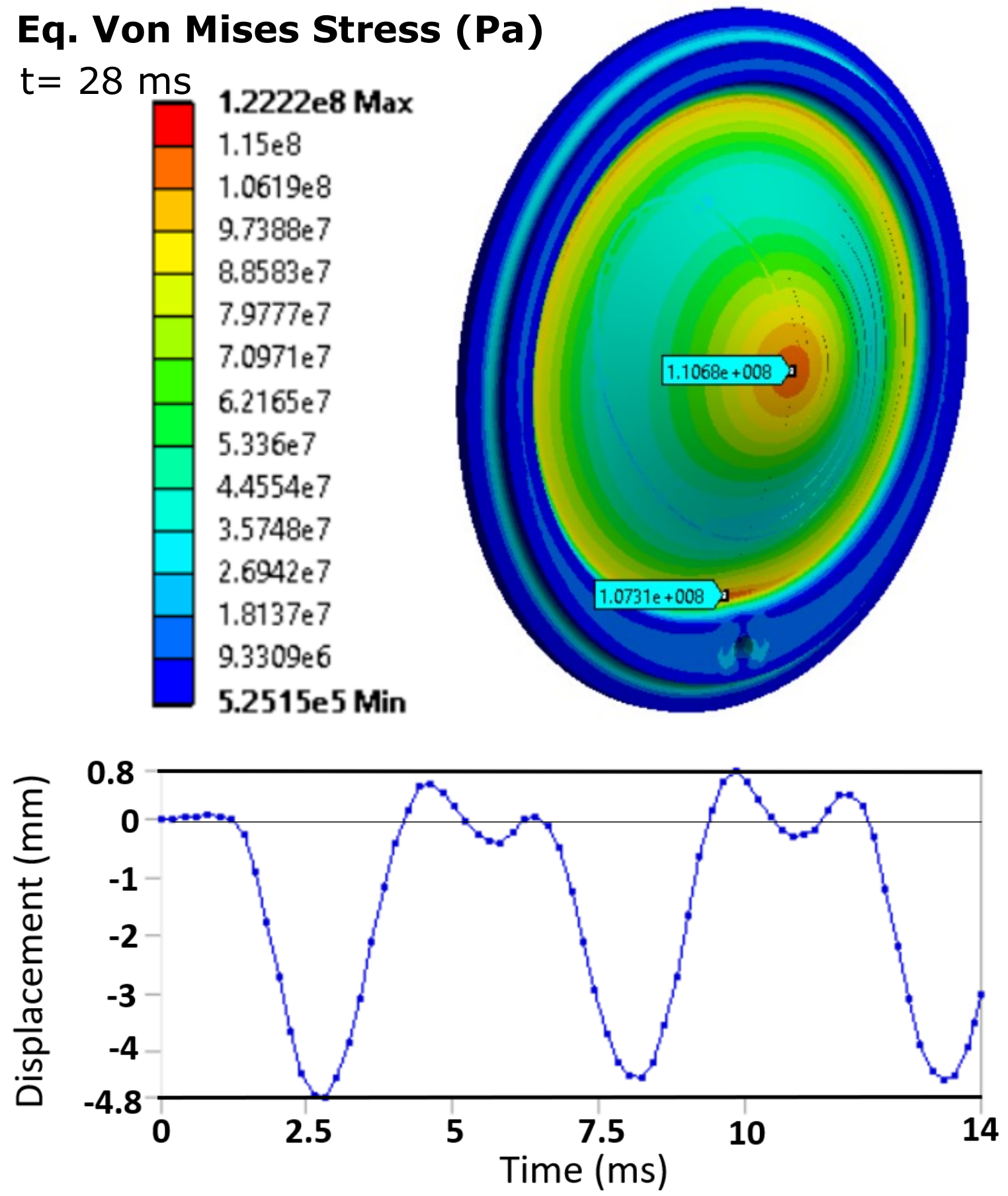}}
\subfigure[]{\includegraphics[width=0.50\linewidth]{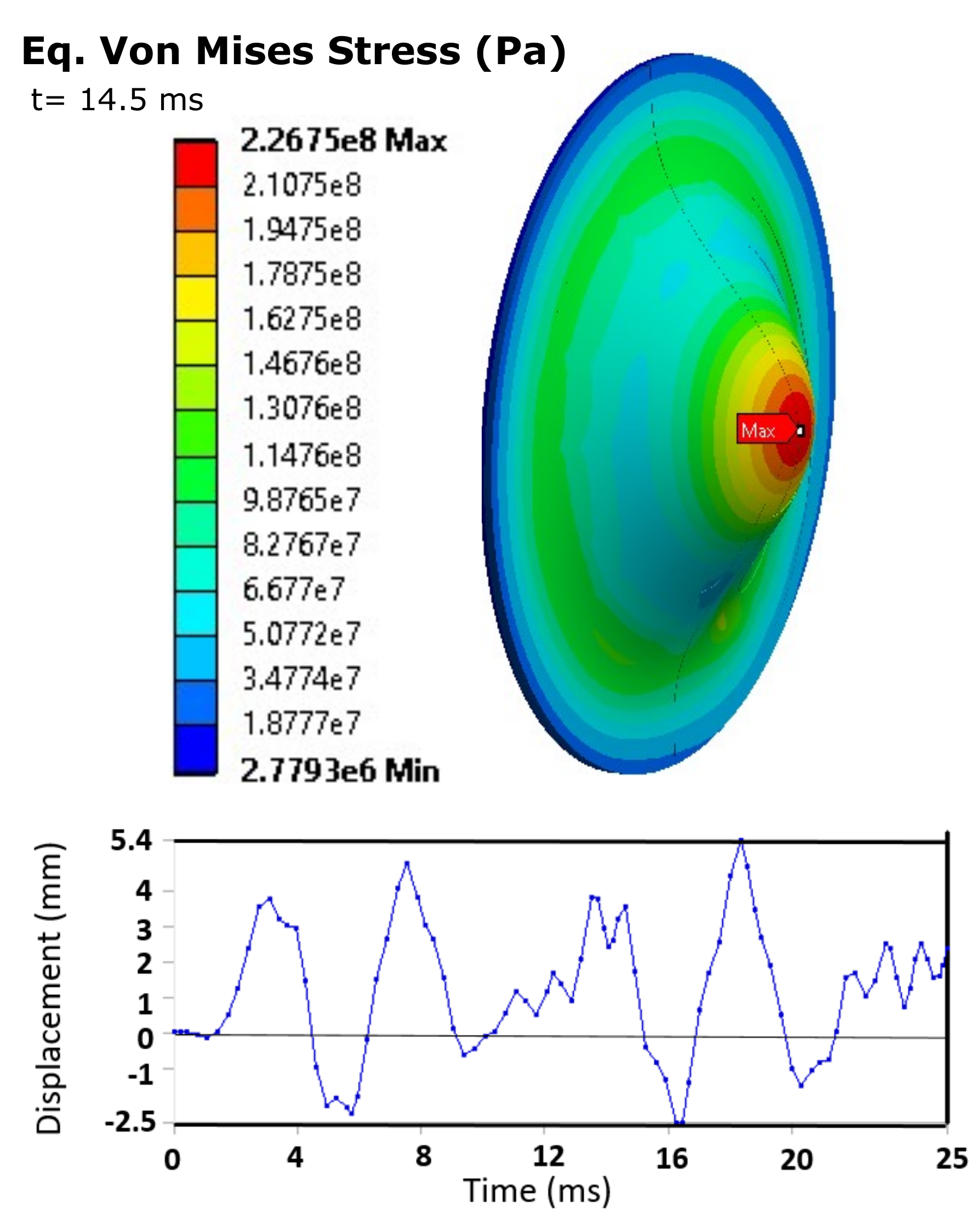}}
\end{center}
 \caption{Contour plots showing the Von Mises stresses in the (a) upstream and (b) downstream window due to the deformation induced by the longitudinal vibrations of the dump, whilst the graphs underneath show the maximum displacements at the center of the upstream and downstream windows, respectively}\label{fig16}
\end{figure}

Fig.~\ref{fig16}a shows the expected maximum Von Mises stress reached in the centre of the upstream window as a consequence of the dump block longitudinal vibrations, simulated as described in Section~\ref{sec:CompuMethod}. The image shows that the maximum Von Mises stresses are in the order of 110 MPa. This maximum is reached 2.8~ms after the beam impact. The second critical region is the connection radius with the thicker part of the window, where the stresses are also of the same order of magnitude. Fig.~\ref{fig16}a-bottom shows the associated displacement of the centre of the window as a function of time. The centre of the window moves up to 4.8~mm in the upstream direction as a consequence of this load. 

 Similarly, Fig.~\ref{fig16}b shows the calculated maximum Von Mises stress reached in the dump downstream window. Maximum Von Mises stresses are of the order of 226~MPa. This maximum is reached 14.5~ms after the beam impact. The centre of the window moves up to 5.4~mm in the upstream direction as a consequence of this load. Maximum stresses and displacements under this load are considerably higher in the dump's downstream window in comparison to the upstream one (even if the displacement load is similar) due to the specific geometry of the dump's downstream window.  

It is worth noting that the stresses associated with this dynamic load, that were not considered in original design studies of the dump~\cite{Evans_2008}, are indeed the dominating ones, for both the upstream and downstream dump windows. In the case of the upstream window, the 110 MPa reached results in a 6.3 safety factor against yielding. For the downstream window, the 226 MPa reached results in a 2.3 safety factor against yielding for the beam dilution failure scenarios in which the highest temperatures are reached (244~$^{\circ}$C).

\subsection{Dump windows fatigue analysis and results}
Following the approach described in Section~\ref{sec:CompuMethod} a multi-axial fatigue assessment was performed for the windows. The assessment was made using the loads of the vessel-induced vibrations, since they are the highest. It results in maximum fatigue equivalent stresses of 140~MPa and 230~MPa for the upstream and downstream windows, respectively. These are well below the fatigue endurance limit in torsion for~$2\cdot10^{6}$ cycles of Ti-Grade 5 at 300~$^{\circ}$C, estimated at 314~MPa.  Hence, we conclude that fatigue is not an issue for the windows for the current parameters of Run~3. 

\subsection{Sealing solution for the upstream window}

As mentioned in Section~\ref{sec:OriginalConf} and \ref{ssec:fast_response}, the metallic Helicoflex\textsuperscript{\textregistered} gaskets used to seal the dump block and connecting line in the original dump design were damaged by the vibration of the dump blocks leading to nitrogen leaks.  It was therefore decided to replace them with elastomeric O-rings for the upstream windows for improved sealing capability under vibration conditions. A second O-ring was used for the connection of the nitrogen supply to each upstream window. 

Based on FLUKA simulations, the O-rings of the upstream windows are expected to absorb a mixed radiation dose of around 0.12~MGy over the Run 3 equipment lifetime. At comparable doses,  modifications of O-ring mechanical properties can be induced, possibly leading to seal failure~\cite{yellow3,davenas,rivaton}. EPDM (ethylene propylene diene monomer) rubber is known to be radiation resistant up to the MGy dose range~\cite{yellow3}. For the upstream windows, O-rings made of EPDM Shieldseal\textregistered~663 produced by James Walker were selected. This is a special grade EPDM-based material designed to be radiation tolerant and declared by the producer as resistant up to a gamma dose of 1.6~MGy~\cite{shieldseal}. Additionally, this material was irradiated in a mixed neutron and gamma radiation field in a previous study~\cite{zenoni2017}, reporting moderate damage at 0.7~MGy and only severe damage at 1.5 MGy~\cite{ThesisMF}, which are well above the  0.12~MGy expected during Run 3. 

\section{Engineering design of the upgraded dump support system}
\label{support_design}

For the Long Shutdown 2 upgrades, it was decided to physically separate the dump blocks from the extraction lines. This change required the development of a new support system for the dump block. Since the dump block's vibrations are intrinsic to its thermo-mechanical behaviour, the aims of the new support system were to allow the dump to vibrate freely, expand and retract without migrating and without transmitting its vibrations or movements to other beam line components. The description of the support system and design analysis are explained in this Section. 

\subsection{Support system description}
\begin{figure}[htbp]
\begin{center}
\subfigure[]{\includegraphics[width=0.75\linewidth]{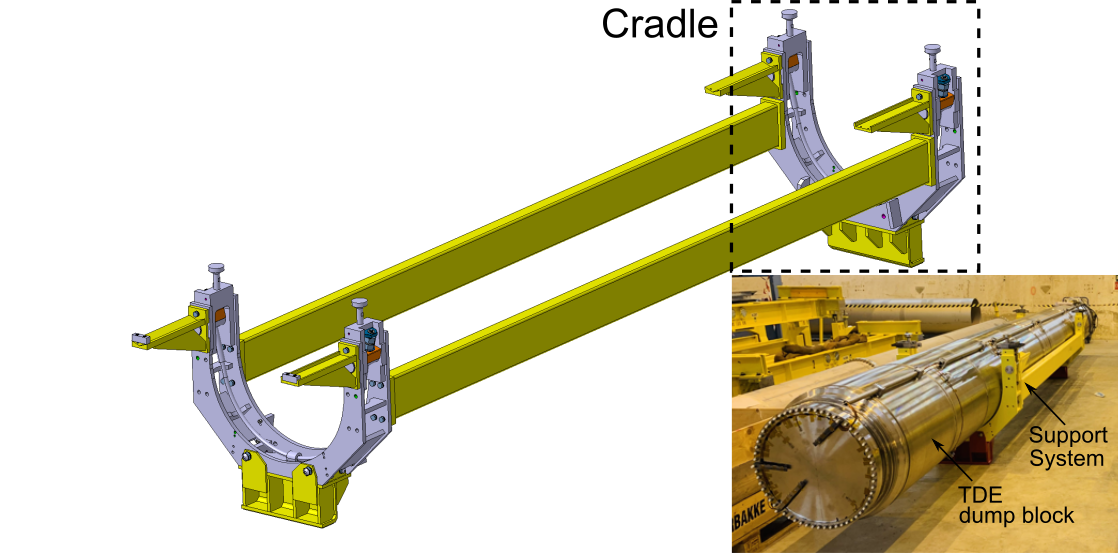}}
\subfigure[]{\includegraphics[width=0.65\linewidth]{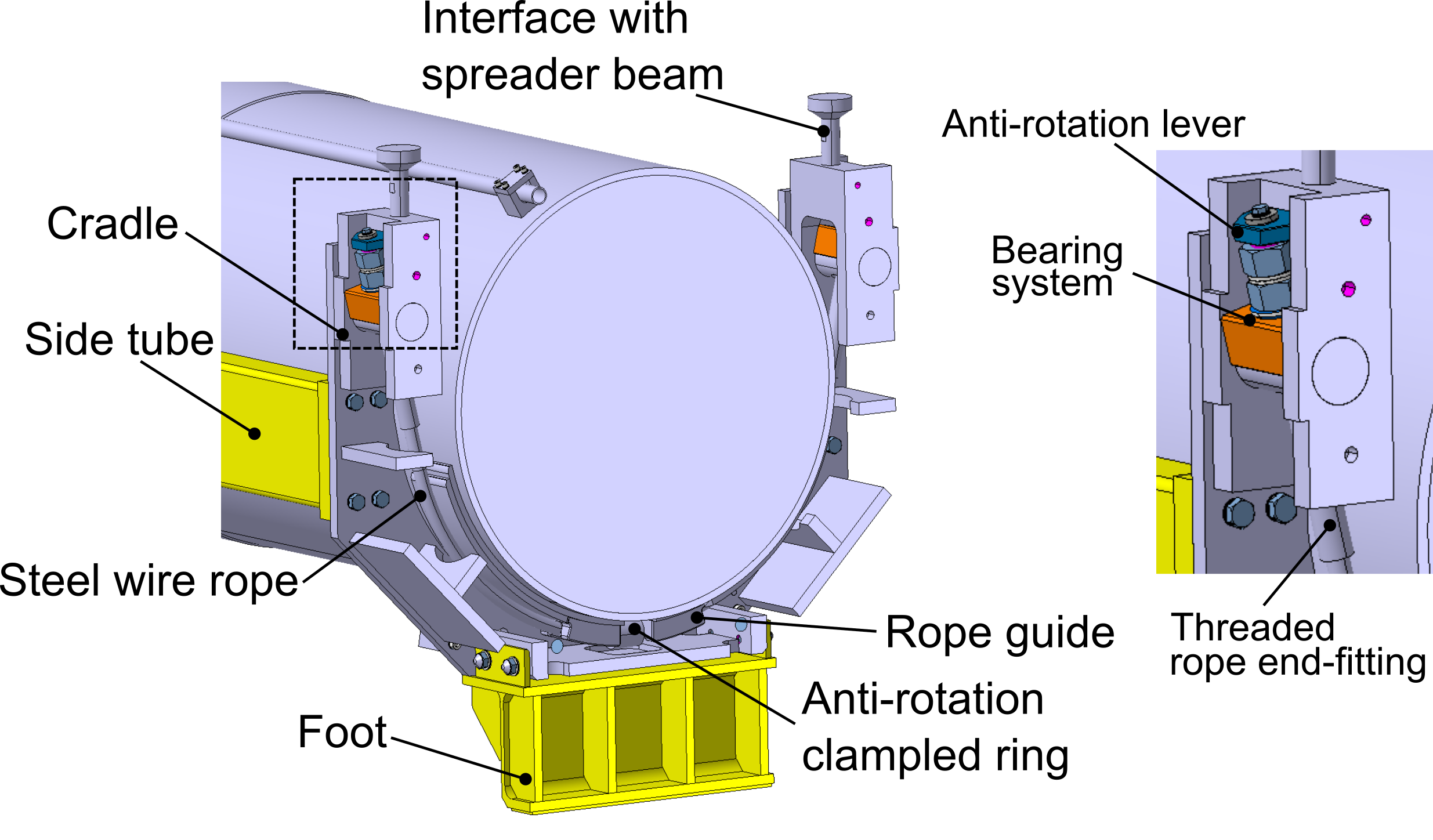}}
\end{center}
 \caption{a) 3D view of the new dump support frame with cradles at each end. The inset photo shows both upgraded dump blocks assembled in their support systems prior to installation. b) Cradle details: the two wire rope loops that support the dump block can be seen inside the cradle structure fitted at each end of the support frame. }\label{fig17}
\end{figure}

A new support system, compatible with the highly radioactive environment, was designed to meet the aims listed above. The system is based on two steel wire rope loops that suspend the dump block inside a frame (see Fig.~\ref{fig17}). This design provides freedom to accommodate vibrations and at the same time avoids any accumulative movement of the dump over repeated expansion and retraction cycles as the dump is brought back to its original position by gravity. The wire ropes are supported at each end by a steel S460ML (1.8838): EN 10025-4-2004 cradle structure that fits around the dump block. The cradles are separated by longitudinal side members (rectangular section tubes in steel EN 10025 S235JR(1.0038) ). Each cradle has a foot that sits on the shielding (the original supports act as guides to position the new support frames). The new support design was designed to fit in the restricted space inside the shielding blocks around the dump block and to be compatible with the remote handling equipment (overhead travelling crane and spreader beam) used for the dump block installation and removal.

\subsection{Dump suspension - steel wire rope selection and assembly  }

The steel wire rope is a key element in the new support system. A $21$ mm diameter galvanized high strength steel wire rope with a configuration 8xK26WS-IWRC, manufactured by DIEPA, was selected for this application (commercial ref: DIEPA H40\textsuperscript{\textregistered}). The two rope assemblies used to support the dump block include swaged end-fittings which interface the ropes to the support frame via spherical bearings. This arrangement is used to accommodate the relative movements between the dump and the support frame. Since the selected rope is not rotation-resistant (ordinary cross-lay configuration), the design incorporates anti-rotation levers at each end of the rope assemblies to prevent unravelling of the wire rope. Details of the cradle can be seen in Fig.~\ref{fig17}b.

\subsection{Thermo-mechanical assessment of the support system frame}
The design of the new support system was developed through an iterative process in parallel with the analysis described above to meet the structural, operational and integration requirements. The new support system does not only have to withstand the dump weight but also its dynamic loads. The dump block and support frame are coupled by the wire ropes and the vibrations from the dump block, although attenuated by ropes, are transmitted through them. In addition, a substantial amount of energy ($4.2$~MJ) is also deposited in the support frame due to the particle shower generated at the beam impact. This is, in fact, the most important load on the frame.

\begin{figure}[htbp]
\begin{center}
\subfigure[][]{\includegraphics[width=0.75\linewidth]{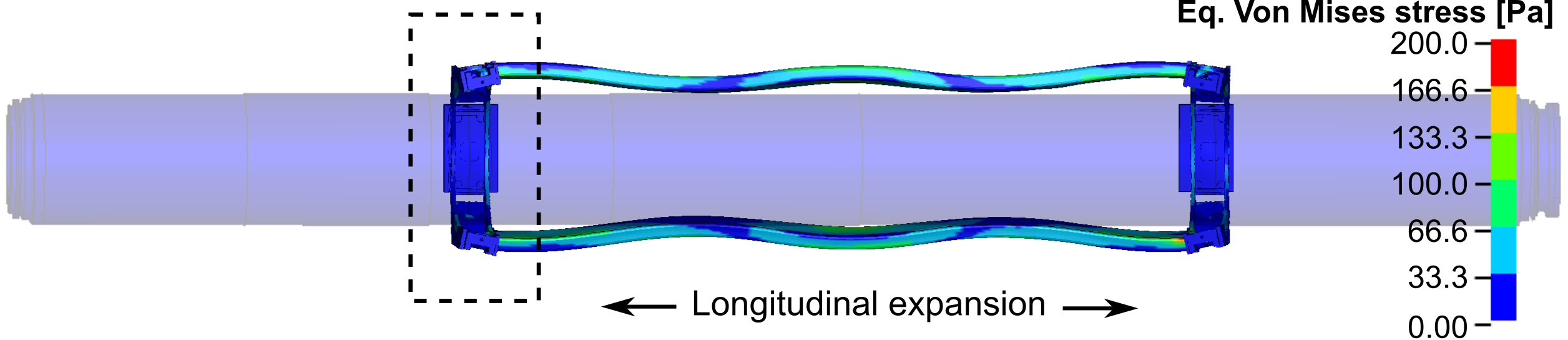}}
\subfigure[][]{\includegraphics[width=0.5\linewidth]{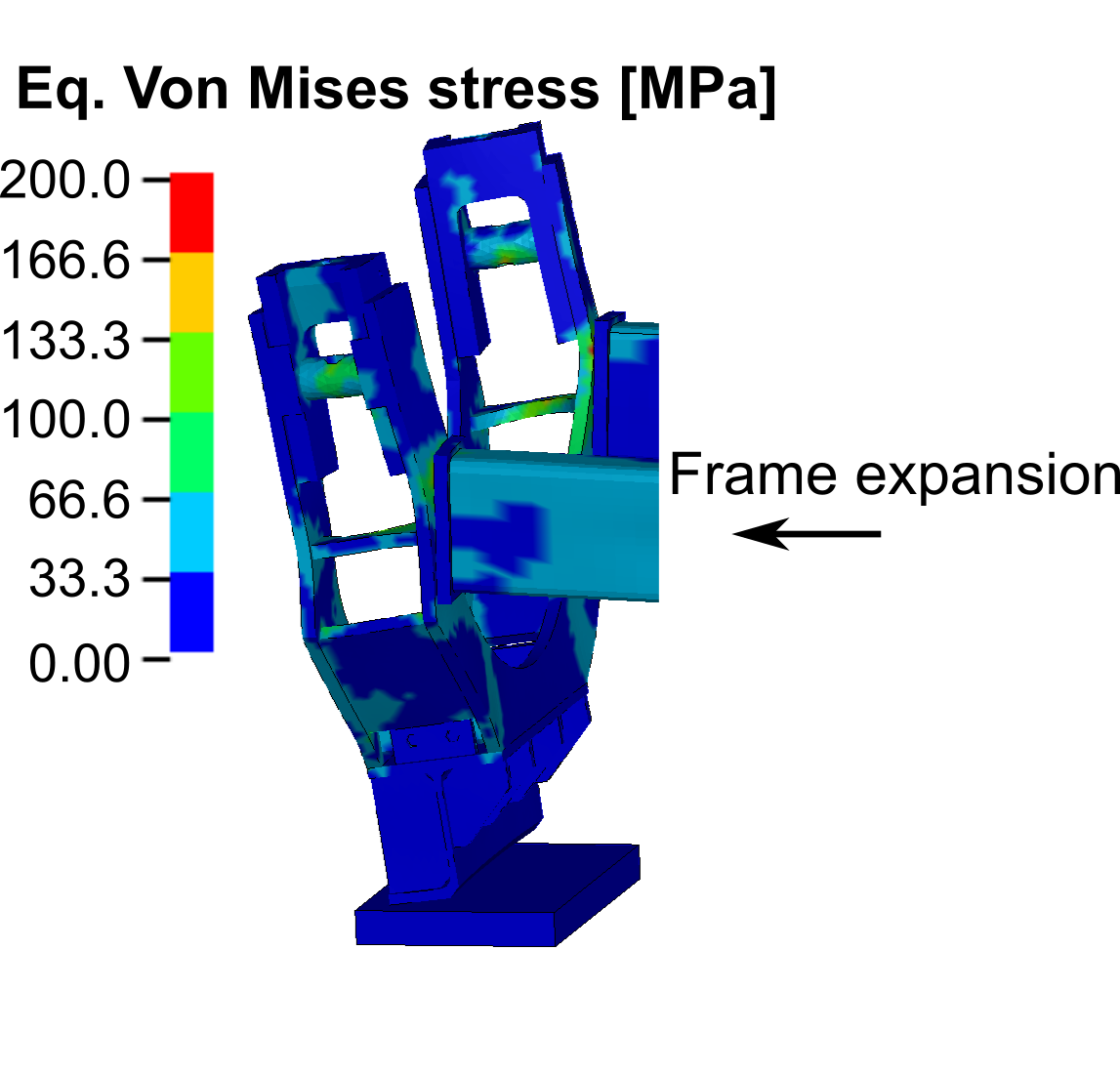}}
\end{center}
\caption{a) Illustration of the mechanical response of the support system frame (viewed from above) after a beam dump event, including b) a detail of the upstream cradle response.}
\label{fig19}
\end{figure}

Fig.~\ref{fig19} illustrates the mechanical response of the support frame. The frame expands longitudinally due to the sudden temperature increase (energy deposition in the side members). As a result, this leads to a global bending deformation of the side tubes and strong dynamic movement of the dump supporting cradles. The design was optimized to keep stresses ($<200$~MPa) under the yield stress of the material ($235$~MPa), reducing at the same time the weight and the inertia effects. According to fatigue analyses, no issues are expected during the support lifetime, and it should be able to withstand more than $35000$ beam dump events.

\subsection{Mechanical assessment of the dump supporting rope}
\label{sec:rope_tests}
\begin{figure}[htbp]
\begin{center}
{\includegraphics[width=0.2\linewidth]{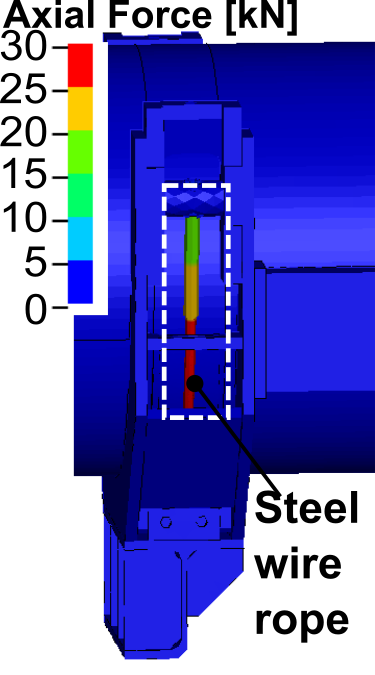}}
{\includegraphics[width=0.65\linewidth]{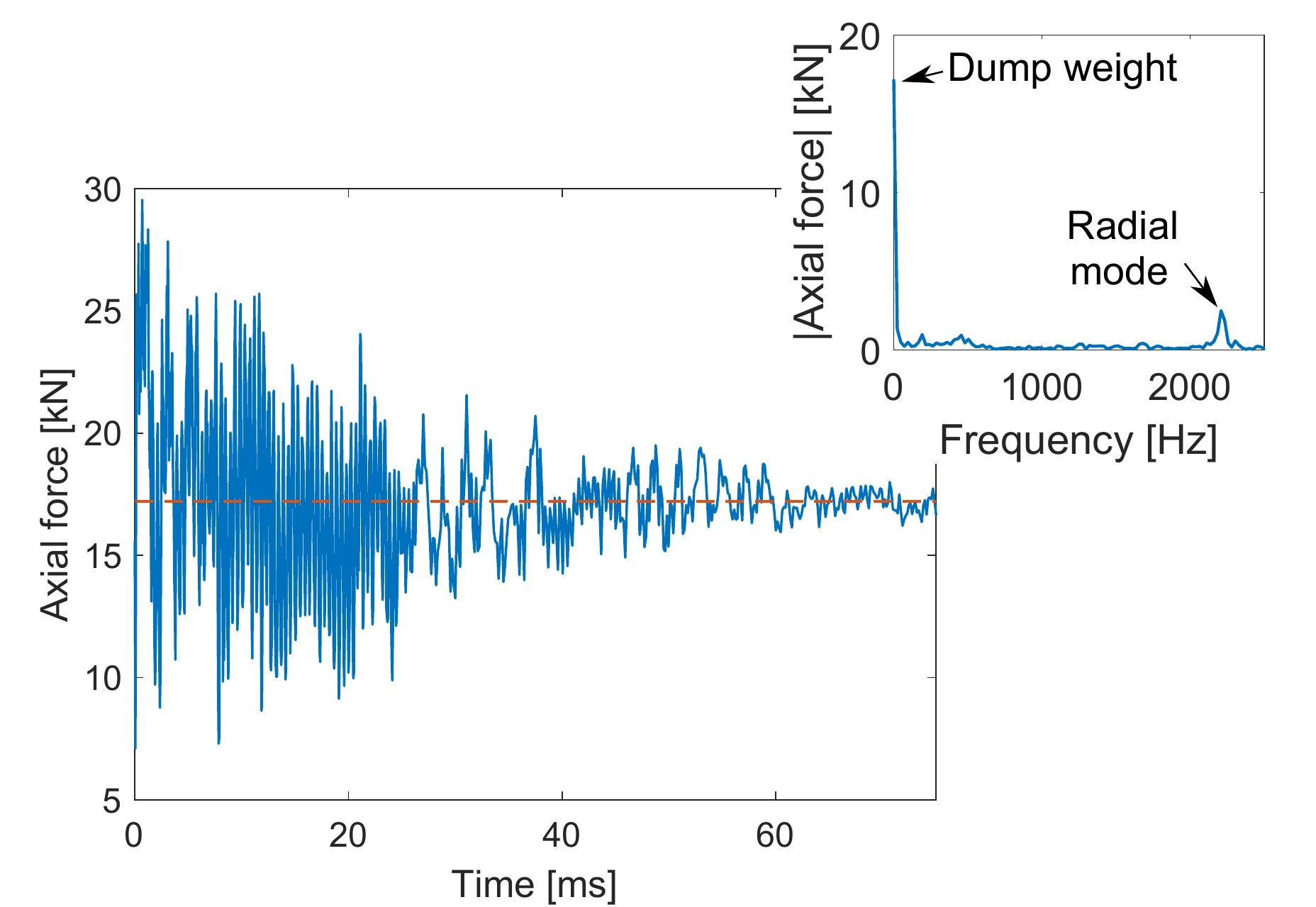}}
\end{center}
\caption{Axial force evolution of the rope after a beam dump event, where the average dashed line corresponds to the weight of the dump block. The rope elongates and contracts at the radial natural frequency of the dump block as shown in the frequency analysis.}
\label{fig20}
\end{figure}

Detailed FEAs were carried out on the supporting ropes, which included the rope-dump interactions (see Fig.~\ref{fig20}). Fig.~\ref{fig20}b shows the axial force in the rope assembly against time, reaching a maximum of roughly $30$~kN. In this application the rope works mainly in tension-tension mode (TT mode) as opposed to a bending mode that is found in lifting equipment where ropes move over pulleys and drums \cite{Ridge2001,feyrer2007}. The rope exhibits oscillations around the static weight of the dump ($17$~kN), and the main frequency matches that of the radial mode of the dump block (see Table~\ref{tab1}, modal analysis measurements). These oscillations tend to attenuate with time due to damping. Considering the rope strength (minimum breaking force of $430$~kN), there is a large quasi-dynamic safety factor ($\approx 13$).

\begin{figure}[htbp]
\begin{center}
{\includegraphics[width=0.6\linewidth]{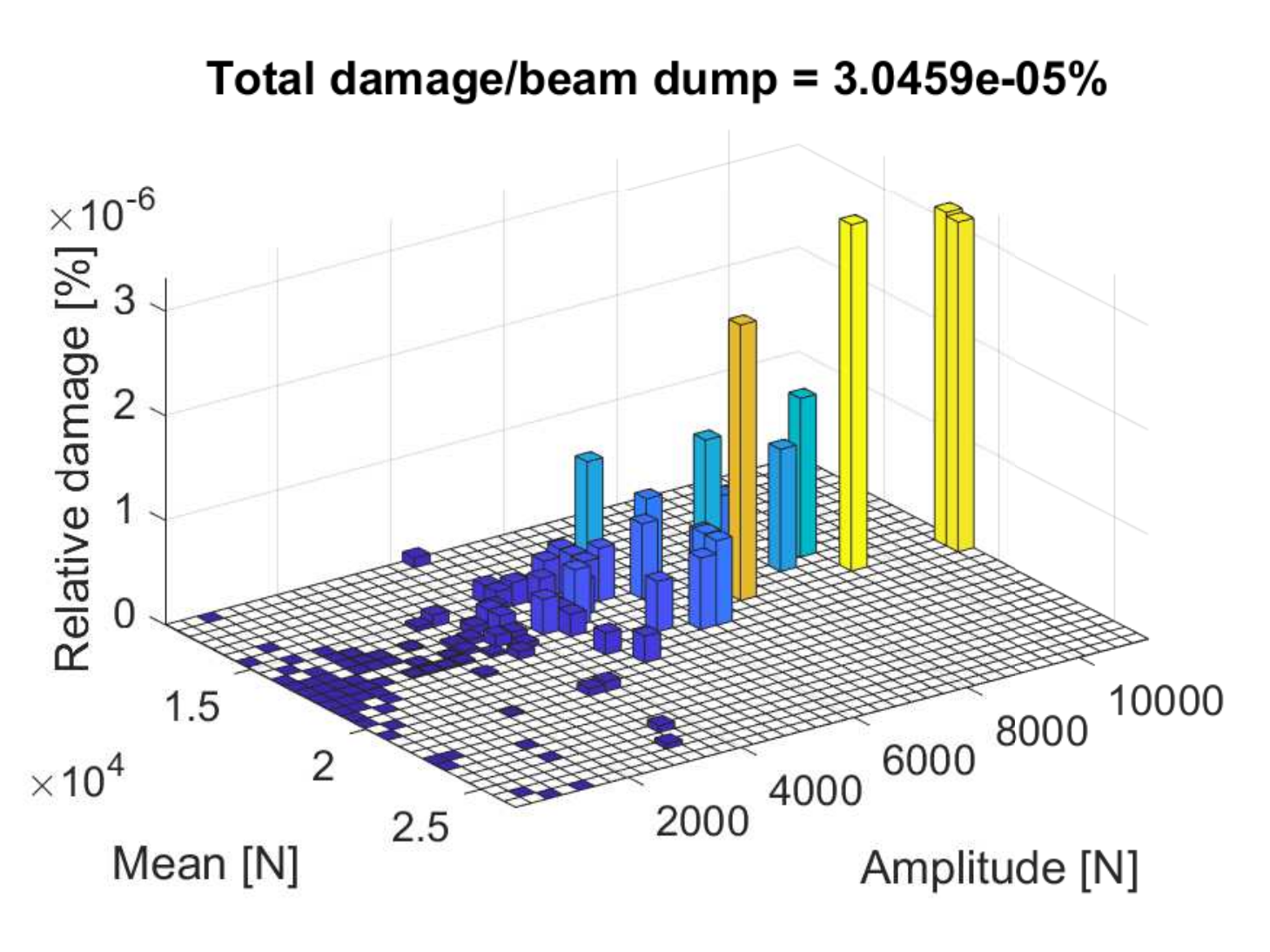}}
\end{center}
\caption{Fatigue damage decomposition of the rope for each beam dump event depending on the mean value and amplitude of the vibrations. Fatigue properties based on literature data~\cite{Ernst2012,feyrer2007}}
\label{fig21}
\end{figure}

As for the other components of the dump block system, the steel wire rope fatigue resistance was assessed. According to literature~\cite{feyrer2007}, fatigue properties of steel wire ropes are mainly driven by the kind of steel and the configuration and structure of the wire rope. M{\"u}ller \cite{muller1966} demonstrates that lubrication has an important effect on rope endurance (e.g. non-lubricated wire ropes have an endurance of about $\approx75\%$ with respect to lubricated ones in TT-fatigue~\cite{muller1966}), hence the importance of grease to service-life. Following the fatigue analyses procedure described in Section~\ref{fatigue_assesment}, a damage of $3\cdot10^{-5}\%$ per beam dump event (where $100\%$ means failure) was estimated (see Fig.~\ref{fig21}), giving a total expected damage after $4$ years of Run~3 operation of $0.05\%$.

A series of tension-tension fatigue tests were carried out to confirm the fatigue performance of the rope assemblies including their end fittings. The tests were performed under constant amplitude, in the range of $[4-30]$~kN, at $2$ Hz on both a fully de-greased (worst case to cover lubricant degradation by radiation) and a greased rope. Samples of degreased rope successfully completed more than $2.24$ million cycles. Inspection showed some minor fretting marks and a subsequent quasi-static tensile test on the de-greased sample resulted in an effective breaking force of $413.7$~kN. As shown in Fig.~\ref{fig21}, only the cycles at high amplitude ($>2$ kN) are considered to produce significant damage, this is, $\approx140$ cycles/beam dump. Therefore, no fatigue issues are expected (safety factor $>10$) during the service life. 

\subsection{Preparation of the dump support ropes for the operating environment}
\begin{figure}[htbp]
\begin{center}
\subfigure[]{\includegraphics[width=0.40\linewidth, trim = 0 1cm 0 0, clip]{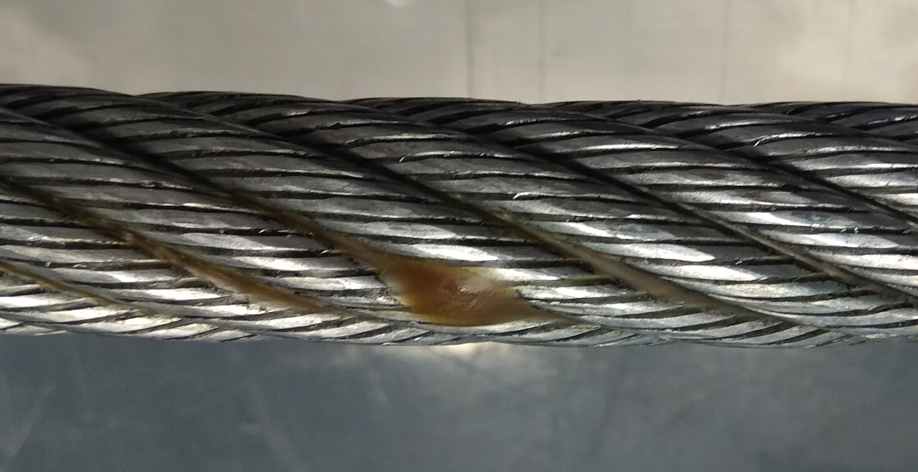}}
\subfigure[]{\includegraphics[width=0.40\linewidth, trim = 0 2cm 0 1.8cm, clip]{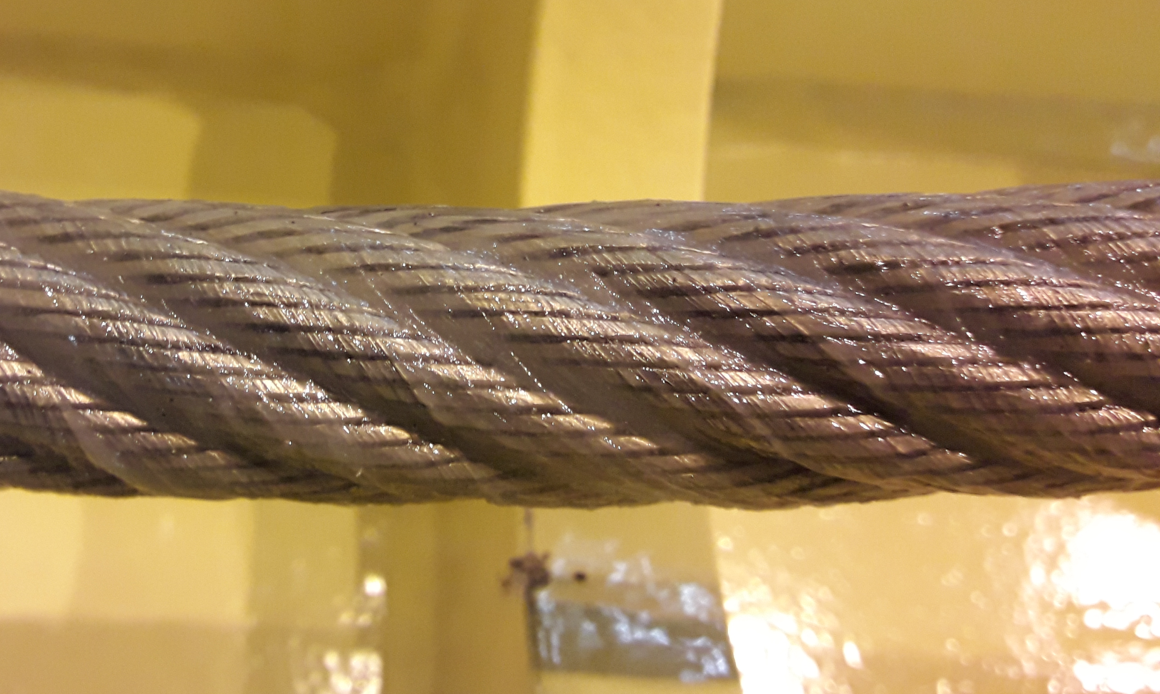}}
\end{center}
 \caption{a) Wire rope with original lubricant (Elaskon\textsuperscript{\textregistered}~SK-DL) showing signs of dropping after exposure to the temperatures expected in mid-Run~3; and b) Beam dump support steel wire rope with a double lubrication: a limited quantity of Elaskon\textsuperscript{\textregistered}~SK-DL between strands and individual wires along with Lubrilog\textsuperscript{\textregistered}~ LX AGFA 2 on the outer surface.}\label{fig18}
\end{figure}

The support system and the ropes are exposed to a radioactive environment and relatively high temperatures. Based on FLUKA-FEA simulations mentioned in Section~\ref{Fluka} and Section~\ref{mechanical_assesment}, the rope is expected to absorb a radiation dose of around $15$ MGy over the equipment lifetime during Run~3 and to operate at temperatures slightly above $100^{\circ}$C. This level of radiation is extremely challenging for polymers and in particular polymeric lubricants\cite{bolt, ferrari2021}, which should therefore ideally be avoided or selected from highly radiation resistant products. Commercial steel wire ropes normally include plastic inserts, whose radiation resistance is unknown and that could be severely damaged at the expected dose levels. To prevent this, the selected rope does not include any plastic inserts. However, as is standard practice during production, the rope is assembled to include a commercial lubricant (Elaskon\textsuperscript{\textregistered}~SK-DL) in order to prevent corrosion and reduce wear. With a drop point around $70$~$^{\circ}$C, this product is expected to melt and partially drop out from the ropes during operation due to the temperatures and vibrations to which it will be exposed (see Fig.~\ref{fig18}a). Moreover, according to the authors' knowledge, the radiation resistance of Elaskon\textsuperscript{\textregistered}~SK-DL has never been experimentally assessed. Considering the available information on its composition and the radiation tolerance of similar products previously tested, it is expected to degrade due to radiation damage before the end of the equipment lifetime. As mentioned above, the fatigue tests on samples of the steel wire rope assemblies were therefore carried out after deep cleaning to remove all lubricant in order to simulate a worst-case scenario of a complete loss of lubricating properties (see Section~\ref{sec:rope_tests}). 

As fatigue safety factors of un-lubricated wire ropes could be increased by using a temperature and radiation resistant lubricant it was decided to investigate suitable alternatives. Several solutions were investigated such as deep cleaning followed by re-greasing with a radiation-tolerant grease, but this cleaning process resulted in a reduced thickness of the galvanised layer on the steel wires. Therefore, a lubrication solution was selected that combines a heat treatment to remove excess Elaskon \textsuperscript{\textregistered}~SK-DL without impacting the galvanised layer, and coating the ropes with a grease having a much higher expected radiation resistance (Fig. ~\ref{fig18}b). Lubrilog\textsuperscript{\textregistered}~LX~AGFA~2 was selected for this application based on its resistance to temperatures up to  $350^{\circ}$C, the high radiation resistance declared by the producer~\cite{lubrilog}, and its Polyphenyl ether-based composition; previous studies~\cite{bolt,jaeri, ferrari2019} demonstrated superior resistance of Polyphenyl ether-based lubricants to radiation. Experimental studies to further assess the radiation tolerance of LX~AGFA~2 are ongoing in collaboration with the producer. 

\section{Conclusions and outlook}
\label{conclusion}
The investigations, analysis and design work carried out to resolve operational problems have led to an improved understanding of the behaviour of the LHC beam dumps in response to the energy deposited by the dumped proton beams. A particularly interesting finding was that the almost instantaneous heating of the dump outer vessel due to the secondary particle shower leads to high frequency and high intensity vibration of the LHC dump block system. Another key finding was that the stresses in the beam windows due to the beam impact on the window itself are dwarfed by the stresses induced by the strong vibrations transmitted from the dump outer vessel to the window.

Although the main focus of the work reported here was to prepare the LHC Beam dumps for the next operational run (Run~3) of the LHC, a secondary goal was to acquire knowledge, develop analysis techniques and gain design, manufacture and operational experience that can be applied to future upgrades of the LHC. In addition to the upgrades of the dumps, supporting systems and beam windows, a new extensive instrumentation system has been installed in order to measure temperatures, strains, vibrations and displacements of the dump during operation. Even though the new instrumentation system is not the main focus of this paper, future publications are planned to report the operational behaviour of the LHC dump blocks and validation of the present analyses, which should be a major input for the design of new dumps to cope with High Luminosity LHC (HL-LHC) energy levels (around 680~MJ/dump).


The process and results presented in this paper are also fully relevant to any future high energy and high intensity hadron or lepton machine, for example FCC~\cite{FCCee,FCChh,HELHC}, ILC or CLIC~\cite{Boland:2210892}, and in general for the design of beam intercepting devices and absorbing systems.

\begin{acknowledgments}
The authors wish to acknowledge the support of CERN's main design office, workshop and mechanical measurement laboratory in the execution of these hardware upgrades and the vital operational support provided by CERN's Radiation Protection, Transport, Robotics, Survey and Coordination teams among others. Special thanks are given to the ACC-CONS (Accelerator Consolidation) Project and CERN Management for their support and trust in the execution of the Long Shutdown 2 upgrades and ensuing activities. The support of the Project Reviewers who provided their time and valuable insights is kindly acknowledged.

\end{acknowledgments}

\bibliography{biblio}

\providecommand{\noopsort}[1]{}\providecommand{\singleletter}[1]{#1}%
\begin{thebibliography}{10}

\bibitem{FlukaWeb}
{FLUKA.CERN} website.
\newblock \url{https://fluka.cern/}.

\bibitem{AMZALLAG1994287}
C.~Amzallag, J.P. Gerey, J.L. Robert, and J.~Bahuaud.
\newblock Standardization of the rainflow counting method for fatigue analysis.
\newblock {\em International Journal of Fatigue}, 16(4):287--293, 1994.

\bibitem{jaeri}
K.~Arakawa and et~al.
\newblock {\em {D}ata on Radiation Resistance of Lubricating Oil - JAERI-M
  87-141}.
\newblock Japan Atomic Energy Research Institute, September 1987.

\bibitem{artoos2007}
Kurt Artoos, Michael Guinchard, Andrea Catinaccio, Keith Kershaw, and Antti
  Onnela.
\newblock Experimental modal analysis of components of the lhc experiments.
\newblock In {\em 2007 IEEE Particle Accelerator Conference (PAC)}, pages
  329--331, 2007.

\bibitem{balthazar2007}
Jos{\'e}~C Balthazar and Lucival Malcher.
\newblock A review on the main approaches for determination of the multiaxial
  high cycle fatigue strength.
\newblock {\em Mechanics of solids in Brazil, Marcilio Alves \& Da Costa
  Mattos}, pages 63--80, 2007.

\bibitem{Battistoni2015}
Giuseppe Battistoni, Till Boehlen, Francesco Cerutti, Pik~Wai Chin,
  Luigi~Salvatore Esposito, Alberto Fassò, Alfredo Ferrari, Anton Lechner,
  Anton Empl, Andrea Mairani, Alessio Mereghetti, Pablo~Garcia Ortega, Johannes
  Ranft, Stefan Roesler, Paola~R. Sala, Vasilis Vlachoudis, and George Smirnov.
\newblock Overview of the {FLUKA} code.
\newblock {\em Annals of Nuclear Energy}, 82:10--18, 2015.
\newblock {Joint} International Conference on Supercomputing in Nuclear
  Applications and Monte Carlo 2013, SNA + MC 2013.

\bibitem{bergman2011}
Theodore~L Bergman, Frank~P Incropera, David~P DeWitt, and Adrienne~S Lavine.
\newblock {\em Fundamentals of heat and mass transfer}.
\newblock John Wiley \& Sons, 2011.

\bibitem{yellow3}
P.~Beynel, P.~Maier, and H.~Schoenbacher.
\newblock {\em {C}ompilation of Radiation Damage Test Data: Materials used
  around high-energy accelerators}.
\newblock European Organisation for Nuclear Research (CERN), yellow report
  cern-82-10, part 3 edition, 1982.

\bibitem{bolt}
R.~O. Bolt and J.~C. Carrol.
\newblock {\em {R}adiation Effects on Organic Materials}.
\newblock Academic Press, 1963.

\bibitem{borrvall2009}
Thomas Borrvall.
\newblock A heuristic attempt to reduce transverse shear locking in fully
  integrated hexahedra with poor aspect ratio.
\newblock In {\em 7th European LS-DYNA Conference, Salzburg}, 2009.

\bibitem{Bruning2004}
Oliver~Sim Brüning, Paul Collier, P~Lebrun, Stephen Myers, Ranko Ostojic, John
  Poole, and Paul Proudlock.
\newblock {\em {LHC Design Report}}.
\newblock CERN Yellow Reports: Monographs. CERN, Geneva, 2004.

\bibitem{Bohlen2014}
T.T. Böhlen, F.~Cerutti, M.P.W. Chin, A.~Fassò, A.~Ferrari, P.G. Ortega,
  A.~Mairani, P.R. Sala, G.~Smirnov, and V.~Vlachoudis.
\newblock The {FLUKA} code: Developments and challenges for high energy and
  medical applications.
\newblock {\em Nuclear Data Sheets}, 120:211--214, 2014.

\bibitem{Cossairt1987}
J~D Cossairt.
\newblock {Review of the abort dump shown in the SSC (superconducting super
  collider) conceptual design report}.
\newblock In {\em Workshop on radiological aspects of SSC operation, Berkeley,
  USA}, 4 1987.

\bibitem{crossland1956}
Bernard Crossland.
\newblock Effect of large hydrostatic pressures on the torsional fatigue
  strength of an alloy steel.
\newblock In {\em Proc. Int. Conf. on Fatigue of Metals}, volume 138, pages
  12--12. Institution of Mechanical Engineers London, 1956.

\bibitem{davenas}
J.~Davenas and et~al.
\newblock {\em {S}tability of polymers under ionising radiation: The many faces
  of radiation interactions with polymers}.
\newblock Nuclear Instruments and Methods in Physics Research Section B: Beam
  Interaction with Materials and Atoms, volume 191, issues 1-4, pages 653-661,
  2002.

\bibitem{Ernst2012}
B.~Ernst.
\newblock {\em The influence of twist on the characteristics of tension-tension
  loaded wire ropes}.
\newblock PhD thesis, University of Stuttgart, 2012.

\bibitem{FCCee}
A.~Abada et~al.
\newblock Fcc-ee: The lepton collider.
\newblock {\em The European Physical Journal Special Topics}, 228(2):261--623,
  2019.

\bibitem{FCChh}
A.~Abada et~al.
\newblock Fcc-hh: The hadron collider.
\newblock {\em The European Physical Journal Special Topics}, 228(4):755--1107,
  2019.

\bibitem{HELHC}
A.~Abada et~al.
\newblock He-lhc: The high-energy large hadron collider.
\newblock {\em The European Physical Journal Special Topics},
  228(5):1109--1382, 2019.

\bibitem{Boland:2210892}
M.~J.~Boland et~al.
\newblock {\em {Updated baseline for a staged Compact Linear Collider}}.
\newblock CERN Yellow Reports: Monographs. CERN, Geneva, Aug 2016.
\newblock Comments: 57 pages, 27 figures, 12 tables.

\bibitem{Evans_2008}
Lyndon Evans and Philip Bryant.
\newblock {LHC} machine.
\newblock {\em Journal of Instrumentation}, 3(08):S08001--S08001, aug 2008.

\bibitem{fatemi1988}
Ali Fatemi and Darrell~F Socie.
\newblock A critical plane approach to multiaxial fatigue damage including
  out-of-phase loading.
\newblock {\em Fatigue \& Fracture of Engineering Materials \& Structures},
  11(3):149--165, 1988.

\bibitem{ThesisMF}
M.~Ferrari.
\newblock {\em {E}xperimental study of radiation resistance in intense neutron
  fields of critical materials and components for the construction of the ESS
  (European Spallation Source) target system}.
\newblock Universit\`a degli Studi di Brescia, Ph.D. Thesis, Cycle XXXII, 2020.

\bibitem{ferrari2019}
M.~Ferrari and et~al.
\newblock {\em {E}xperimental study of consistency degradation of different
  greases in mixed neutron and gamma radiation}.
\newblock Heliyon vol.5, Issue~9, e02489, 2019.

\bibitem{ferrari2021}
M.~Ferrari and et~al.
\newblock {\em {S}election of radiation tolerant commercial greases for
  high-radiation areas at CERN: methodology and applications}.
\newblock Submitted for publication in Nuclear Materials and Energy, 2021.

\bibitem{feyrer2007}
Klaus Feyrer.
\newblock {\em {Wire ropes. Tension, Endurance, Reliability}}.
\newblock Springer, 2007.

\bibitem{goddard2003}
B~Goddard, M~Gyr, J~Uythoven, R~Veness, and W~Weterings.
\newblock Lhc beam dumping system: Extraction channel layout and acceptance.
\newblock In {\em Proceedings of the 2003 Particle Accelerator Conference},
  volume~3, pages 1646--1648. IEEE, 2003.

\bibitem{Guinchard2018}
M.~Guinchard et~al.
\newblock Experimental modal analysis of lightweight structures used in
  particle detectors: Optical non-contact method.
\newblock In {\em Proc. 9th Int. Particle Accelerator Conf. (IPAC'18)}, pages
  2565--2567. JACoW Publishing, Apr.-May 2018.
\newblock https://doi.org/10.18429/JACoW-IPAC2018-WEPMF079.

\bibitem{Hanna1991}
B.~M. Hanna and C.~K. Crawford.
\newblock Construction of a new tevatron collider beam abort dump.
\newblock {\em Conference Record of the 1991 IEEE Particle Accelerator
  Conference}, pages 970--972 vol.2, 1991.

\bibitem{henderson2009}
Mark Henderson.
\newblock {Scientists cheer as protons complete first circuit of Large Hadron
  Collider}.
\newblock {\em Times Online}, 2008.

\bibitem{Karastathis2019}
N~Karastathis et~al.
\newblock {LHC Run 3 Configuration Working Group Report}.
\newblock In {\em Proceedings of the 9th Evian Workshop on LHC Beam Operation,
  Evian Les Bains, France}, pages 273--284, Geneva, Switzerland, 2019.

\bibitem{Kidd1981}
John~M. Kidd, Nikolai~V. Mokhov, T.~Murphy, Mark~A. Palmer, Timothy~Edward
  Toohig, Frank Turkot, and A.~VanGinneken.
\newblock A high intensity beam dump for the tevatron beam abort system.
\newblock {\em IEEE Transactions on Nuclear Science}, 28:2774--2776, 1981.

\bibitem{Lamont2013}
Mike Lamont.
\newblock {The first years of LHC operation for luminosity production}.
\newblock In {\em Proceedings of the 4th International Particle Accelerator
  Conference, MOYAB101, Shanghai, China}, pages 6--10, 2013.

\bibitem{liu2008}
Yucheng Liu.
\newblock Ansys and ls-dyna used for structural analysis.
\newblock {\em International Journal of Computer Aided Engineering and
  Technology}, 1(1):31--44, 2008.

\bibitem{lubrilog}
LUBRILOG\textsuperscript{\textregistered}.
\newblock {\em {T}echnical datasheet of
  Lubrilog\textsuperscript{\textregistered} LX AGFA 2}.
\newblock 2013.

\bibitem{madhusudana1996}
Chakravarti~V Madhusudana and CV~Madhusudana.
\newblock {\em Thermal contact conductance}, volume~79.
\newblock Springer, 1996.

\bibitem{maestre2021}
J.~Maestre and et~al.
\newblock {Sigraflex® studies for LHC CERN beam dump: Summary and
  Perspective}.
\newblock In {\em 12th International Particle Accelerator Conf. (IPAC’21),
  Campina, Brazil}, 2021.

\bibitem{Majidi2008}
B.~Majidi.
\newblock Fatigue life and short crack behavior in ti-6al-4v alloy;
  interactions of foreign object damage, stress, and temperature.
\newblock {\em Metallurgical and Materials Transactions A}, 39:772–777, 2008.

\bibitem{martin2021}
J.M. Martin~Ruiz and et~al.
\newblock {\em {P}ractical challenges of the LHC Main Beam Dump upgrades}.
\newblock CERN. Geneva. ATS Department, 2021.
\newblock CERN-ACC-NOTE-2021-0027.

\bibitem{miner1945}
M.~A. Miner.
\newblock Cumulative damage in fatigue.
\newblock {\em Journal of Appl. Mech.}, 12:159--164, 1945.

\bibitem{muller1966}
H~M{\"u}ller.
\newblock {Drahtseile im Kranbau, Auswahl und Betriebsverhalten}.
\newblock {\em VDI-Berichte}, (98):714--716, 1966.

\bibitem{narvydas2014}
E~Narvydas and N~Puodziuniene.
\newblock Applications of sub-modeling in structural mechanics.
\newblock In {\em Proceedings of 19th International Conference. Mechanika,
  Kaunas, Lithuania}, pages 172--6, 2014.

\bibitem{Papaphilippou2014}
Y.~Papaphilippou, H.~Bartosik, G.~Rumolo, and D.~Manglunki.
\newblock {Operational beams for the LHC}.
\newblock In {\em Proceedings of Chamonix 2014 Workshop on LHC Performance,
  Chammonix, France}, pages 80--84, Geneva, Switzerland, 2014.

\bibitem{Ridge2001}
I.M.L Ridge, C.R Chaplin, and J~Zheng.
\newblock Effect of degradation and impaired quality on wire rope bending over
  sheave fatigue endurance.
\newblock {\em Engineering Failure Analysis}, 8(2):173--187, 2001.

\bibitem{rivaton}
A.~Rivaton, Cambon S., and Gardette J.-L.
\newblock {\em {R}adiochemical ageing of EPDM elastomers.: 2. Identification
  and quantification of chemical changes in EPDM and EPR films
  $\gamma$-irradiated under oxygen atmosphere}.
\newblock Nuclear Instruments and Methods in Physics Research Section B: Beam
  Interaction with Materials and Atoms, volume 227, Issues 3, pages 343-356,
  2005.

\bibitem{Rossi2011}
L.~Rossi.
\newblock {LHC Upgrade Plans: Options and Strategy}.
\newblock In {\em {I2nd International Particle Accelerator Conference. San
  Sebastian, Spain}}, 2011.

\bibitem{schmidt2016high}
Burkhard Schmidt.
\newblock The high-luminosity upgrade of the lhc: Physics and technology
  challenges for the accelerator and the experiments.
\newblock In {\em Journal of Physics: Conference Series}, volume 706, page
  022002. IOP Publishing, 2016.

\bibitem{Scislo2020}
Lukasz Scislo.
\newblock {Experimental modal analysis of LHC beam dump on the final support
  for run III}.
\newblock Technical report, CERN. EDMS no.: Report 2339554, 2020.

\bibitem{Silva2012}
J{\'u}lio M.~M. Silva and Nuno M.~M. Maia.
\newblock {\em Modal analysis and testing}, volume 363.
\newblock Springer Science \& Business Media, 2012.

\bibitem{sines1959}
George Sines, Joseph~Lewis Waisman, and Thomas~J Dolan.
\newblock {\em Metal Fatigue [by] Thomas J. Dolan [and Others] Edited by George
  Sines and JL Waisman}.
\newblock McGraw-Hill, 1959.

\bibitem{Steerenberg2019}
R.~Steerenberg et~al.
\newblock {O}peration and {P}erformance of the {C}ern {L}arge {H}adron
  {C}ollider {D}uring {P}roton {R}un 2.
\newblock In {\em Proc. 10th International Particle Accelerator Conference
  (IPAC'19), Melbourne, Australia, 19-24 May 2019}, pages 504--507, Geneva,
  Switzerland, 2019. JACoW Publishing.

\bibitem{sun2000}
JS~Sun, KH~Lee, and HP~Lee.
\newblock Comparison of implicit and explicit finite element methods for
  dynamic problems.
\newblock {\em Journal of materials processing technology}, 105(1-2):110--118,
  2000.

\bibitem{tirelli2010}
Daniel Tirelli.
\newblock Modal analysis of small \& medium structures by fast impact hammer
  testing method.
\newblock Technical report, Tech. Rep. EUR, 24964 EN, Joint Research Centre,
  Publications Office of the~…, 2010.

\bibitem{van1989}
Dang Van~K, B~Griveau, MW~Brown, K~Miller, et~al.
\newblock On a new multiaxial fatigue limit criterion: theory and application.
\newblock 1989.

\bibitem{shieldseal}
James Walker\textregistered.
\newblock {\em {M}aterials for use in nuclear applications}.
\newblock Issue number 2, BP4793 0419/200, CPN000087082, 2019.

\bibitem{wozniak:icalepcs2019-wepha163}
J.P. Wozniak and C.~Roderick.
\newblock {NXCALS - Architecture and Challenges of the Next CERN Accelerator
  Logging Service}.
\newblock In {\em Proc. ICALEPCS'19}, number~17 in International Conference on
  Accelerator and Large Experimental Physics Control Systems, pages 1465--1469.
  JACoW Publishing, Geneva, Switzerland, 08 2020.
\newblock https://doi.org/10.18429/JACoW-ICALEPCS2019-WEPHA163.

\bibitem{Zazula1996er}
J.~Zazula and S.~Peraire.
\newblock {LHC beam dump design study: Part 1. Simulation of energy deposition
  by particle cascades: Implications for the dump core and beam sweeping
  system}.
\newblock 10 1996.

\bibitem{Zazula1996nd}
J.~Zazula and S.~Peraire.
\newblock {LHC Beam Dump Design Study, Part 2}: {T}hermal analysis; implication
  for abort repetition and cooling system.
\newblock 12 1996.

\bibitem{zazula1997}
J.~Zazula and S.~Peraire.
\newblock {Design studies of the LHC beam dump}.
\newblock In {\em 3rd Workshop on Simulating Accelerator Radiation
  Environments, Tsukuba, Japan,}, pages 150--164, May 1997.

\bibitem{zenoni2017}
A.~Zenoni and et~al.
\newblock {\em {R}adiation resistance of elastomeric O-rings in mixed neutron
  and gamma fields: Testing methodology and experimental results}.
\newblock Review of Scientific Instruments 88, 113304, 2017.

\end{thebibliography}
\bibliographystyle{plain}

\end{document}